%% file: main.tex
\newif\ifconference
\conferencetrue
\newif\ifClassicalAlgs
\ClassicalAlgstrue
\documentclass{article}
\usepackage[a4paper, total={6.6in, 10in}, bottom=1in]{geometry}

\usepackage{amsmath}
\usepackage{amsfonts}
\usepackage{dsfont}
\usepackage{mathtools}
\usepackage{graphicx}
\usepackage{url}
\usepackage{color}
\usepackage{wrapfig}
\usepackage{float} 
\usepackage{siunitx}
\usepackage[table]{xcolor}
\usepackage[multiple]{footmisc}
\usepackage{qcircuit}
\usepackage[boxruled,ruled,vlined]{algorithm2e}
\usepackage{soul}
\usepackage{multirow}

\usepackage{pgfplots}
\usepackage{pgfplotstable}
\usepackage{tikz} 
\pgfplotsset{compat=1.17}

\usepackage{appendix}
\usepackage{authblk}

\newcommand{\e}{\mathrm{e}}
\renewcommand{\i}{\mathrm{i}}
\newcommand{\NN}{\mathbb{N}}
\newcommand{\RR}{\mathbb{R}}
\newcommand{\ZZ}{\mathbb{Z}}

\newcommand{\ket}[1]{|#1\rangle}
\newcommand{\bra}[1]{\langle #1 |}

\newcommand{\E}{\mathbb{E}}

\newcommand{\LG}{\textrm{LG}}
\newcommand{\VG}{\textrm{VG}}
\newcommand{\SA}{\textrm{SA}}
\newcommand{\GSA}{\textrm{GSA}}
\newcommand{\xQAOA}{xQAOA\xspace}

\newcommand{\Xgate}{\mathsf{X}}
\newcommand{\Ygate}{\mathsf{Y}}
\newcommand{\Zgate}{\mathsf{Z}}
\newcommand{\Rgate}{\mathsf{R}}
\newcommand{\Hourglass}{{\mathrlap{\Zgate}\Xgate}}

\newcommand{\Cop}{\mathsf{Cop}}
\newcommand{\CopEn}{\mathsf{Cop}^e_n}
\newcommand{\CopOn}{\mathsf{Cop}^o_n}
\newcommand{\Copn}{\mathsf{Cop}_n}
\newcommand{\QKP}{\textsc{QKP}}

\colorlet{LGcolor}{blue!80!green!20}
\colorlet{VGcolor}{blue!50!green!50}
\colorlet{SAcolor}{pink}
\colorlet{GSAcolor}{violet!50!}

\colorlet{HGcolor}{orange}
\colorlet{Copcolor}{yellow}
\colorlet{Xcolor}{brown}

\usepackage{etoolbox}
\patchcmd{\thebibliography}{\section*{\refname}}{}{}{}

\usepackage[english]{babel}

\title{Quantum Optimization Heuristics with an Application to Knapsack Problems}

\author[1, 2]{Wim van Dam}
\author[3]{Karim Eldefrawy}
\author[3]{Nicholas Genise}
\author[4, 2]{Natalie Parham}
\affil[1]{UC Santa Barbara}
\affil[2]{QC Ware Corp.}
\affil[3]{SRI International}
\affil[4]{Institute for Quantum Computing, University of Waterloo}

\begin{document}

\maketitle
\thispagestyle{plain}
\pagestyle{plain}
\begin{abstract}
This paper introduces two techniques that make the standard Quantum Approximate Optimization Algorithm (QAOA) more suitable for constrained optimization problems. 
The first technique describes how to use the outcome of a prior greedy classical algorithm to define an initial quantum state and mixing operation to adjust the quantum optimization algorithm to explore the possible answers around this initial greedy solution. 
The second technique is used to nudge the quantum exploration to avoid the local minima around the greedy solutions. 
To analyze the benefits of these two techniques we run the quantum algorithm on known hard instances of the Knapsack Problem using unit depth quantum circuits. 
The results show that the adjusted quantum optimization heuristics typically perform better than various classical heuristics. 

\end{abstract}


\input{Introduction.tex}
\input{QAOA_and_Variants.tex}

\input{Knapsack_Background.tex}
\input{QAOA_for_KP}

\input{Experiments_and_Results.tex}

\section{References}
\bibliographystyle{IEEEtran}
\bibliography{refs}

\input{appendix}
\end{document}

%% file: Introduction.tex
\section{Introduction}

Constrained optimization involves maximizing (or minimizing) an objective function in some variables while satisfying constraints on those variables. The objective function is typically a cost function (sometimes called energy function), often to be minimized, or a reward function (sometimes called utility functions) to be maximized. 

In this paper, we focus on the Knapsack Problem (KP) as a representative of a class of combinatorial optimization which appear in real-world decision-making processes in a wide variety of fields. 
Informally, in an instance of KP, one is given a set of items, each with a weight and a value, and the objective is to determine how many items from each type to include in a collection resulting in a total weight less than or equal to a given limit (constraint) and the total value (objective function) is maximized. 
The KP has been studied for more than a century, with early works dating as far back as 1897.  The name ``knapsack'' denotes a problem faced by someone who is constrained by a fixed-size knapsack and must fill it with valuable items.
The problem often arises in resource allocation where the decision-makers have to choose from a set of non-divisible projects or tasks under a fixed budget or time constraint, respectively.
Other examples of applications can be found in manufacturing during the process of attempting to find the least wasteful way to cut raw materials. 
Another example is in finance when selecting investments and portfolios, or selecting assets for asset-backed securities.

\subsection{Contributions}

This  paper  introduces  new  techniques  that  make the  standard  QAOA  more  suitable  for  constrained  optimization  problems. 
The  first  technique  describes  how  to  use  the  outcome  of  a prior  greedy  classical  algorithm  to  define  an  initial  quantum state  and  mixing  operation  to  adjust  the  quantum  optimization algorithm  to  explore  the  possible  answers  around  this  initial greedy  solution.  
The  second  technique  is  used  to  nudge  the quantum exploration to avoid the local minima around the greedy solutions.  
To  analyze  the  benefits  of  these  two  techniques  we run  the  quantum  algorithm  on  known  hard  instances  of  the KP  using  unit  depth  quantum  circuits.  The results  show  that  the  adjusted  quantum  optimization  heuristics typically  perform  better  than  various  classical  heuristics. 
It should be noted that our first technique, when applied to problems with a linear objective function (as is the case with KP), can be simulated by a classical circuit with the same complexity as our quantum circuit, and thus can be interpreted as a \emph{quantum-inspired} classical algorithm.

\subsection{Outline of Paper}
The rest of the manuscript is organized as follows. 
Section~\ref{section:QAOA} briefly describes prior work on the Quantum Approximate Optimization Algorithm (QAOA).  
In Section~\ref{section:xQAOA} we describe our main contributions, which are two extensions of QAOA to better deal with constrained optimization problems. 
In Section~\ref{section:KP}  we introduce definitions of the Knapsack Problem (KP) and  discuss its hardness. 
Section~\ref{section:QAOAforKP} discusses limitations of the standard QAOA for solving the KP, after which we describe our proposed xQAOA variations. 
In Section~\ref{section:results} we present our experimental results on the performance of xQAOA on hard KP Instances, and we compare this performance with that of several classical optimization algorithms. 
Section \ref{sec:concusion} concludes the paper with a discussion of future work.

%% file: QAOA_and_Variants.tex
\section{Quantum Approximate Optimization}\label{section:QAOA}

\subsection{Combinatorial Optimization}\label{ssect:comb}
	
	Combinatorial optimization problems encompass a large
	family of search problems on a finite set $S$
	where one must find an optimal solution, measured
	by some (efficiently computable) value function
	$v: S \rightarrow \RR$ such that the solution
	is in some efficiently recognizable feasible set,
	$F \subseteq S$.
	In other words, find 
		$x^* = \arg\max_{x\in F \subseteq S} v(x)$.
    Without loss of generality,
	we assume that values are always positive.
	
The Quantum Approximate Optimization Algorithm (QAOA) of Farhi et al. \cite{farhi_etal-2014}, 
can be viewed as a discrete, or bang-bang, variant of quantum adiabatic optimization.  
In this study, we extended the standard QAOA approach and tailor it to the Knapsack Problem. 

\subsection{QAOA, sensu stricto}
As defined by Farhi et al. \cite{farhi_etal-2014}, the quantum circuit for QAOA works as follows. 
In a typical QAOA of depth $p$ one searches for the minimum of a cost function $C:\{0,1\}^n\rightarrow \ZZ$ by applying a sequence 
\begin{align}
U^{B}(\beta_p)
U^{C}(\gamma_p)
\cdots 
U^{B}(\beta_1)
U^{C}(\gamma_1)
\ket{\Psi_0}
= \ket{\Psi^C_p(\beta,\gamma)}
\end{align}
of $2p$ unitary transformations on a standard initial state $\Psi_0$, which is the uniform superposition over all $n$-bit strings. 
For each level $1\leq j \leq p$ there are two angles $(\beta_j,\gamma_j)\in \RR^2$ that determine the unitaries 
\begin{align}
U^B(\beta_j) = \exp(-i\beta_j B) \text{~and~}
U^C(\gamma_j) = \exp(-i\gamma_j C)
\end{align}
such that for all $x\in\{0,1\}^n$:
\begin{align}
\begin{cases}
U^B(\beta_j)\ket{x} = e^{-i\beta_j \sum_{i=1}^n X_i}\ket{x},\\
U^C(\gamma_j)\ket{x} = e^{-i\gamma_j C(x)}\ket{x}.
\end{cases}
\end{align}
Note that $B$ and $C$ are $n$-qubit Hamiltonians with 
\begin{align}
B & = \sum_{i=1}^n \Xgate_i, & 
C  & = \sum_{x\in\{0,1\}^n}C(x)\ket{x}\bra{x}.
\end{align}
The $U^B(\beta_j)$ transformations thus induces a $C$-independent 
quantum random walk along the edges of the hypercube $\{0,1\}^n$, while $U^C(\gamma_j)$ applies a phase rotation proportional to the cost $C$. 
We refer to $B$ as the \emph{mixing Hamiltonian}, and to $C$ as the \emph{energy function.} 

The goal is to arrange all of this in such a way that the final state $\Psi^C_p$ is concentrated on those strings $x$ that optimize $C$. 
To tune the QAOA for a class of problem instances $\mathcal{C} = \{C\}$, one decides on a depth $p$ and $2p$ angles $(\beta,\gamma)\in \RR^{2p}$ such that for each problem instance $C$, the final state $\Psi^C_p$ is as optimal as possible.  
In the past five years, it has become clear that QAOA is a promising way of attacking unconstrained binary optimization problems where the cost function $C$ is a polynomial-sized sum of bit-to-bit interactions.  

\subsection{Penalty Functions to Enforce Constraints}\label{sec:penalty_function_approach}
To deal with constraints on the allowed $x$ in an optimization problem a standard technique is to add a penalty function to the objective function, thereby reducing it to an unconstrained problem. 
Hence, for example, the minimization of an $n$-bit function $f:\{0,1\}^n\rightarrow \RR$ under the Hamming weight constraint $x_1 + \cdots +x_n\leq n/2$, is rephrased as the minimization over all $n$-bit strings of the adjusted function $f(x)+p(x)$, where $p(x)=0$ for all allowed  $x_1+\cdots +x_n\leq n/2$, while $p(x)$ is large for the disallowed $x_1+\cdots +x_n>n/2$. 
By picking the penalty large enough we can force the fact that the unconstrained minimum of $f(x)+p(x)$ coincides with the constrained minimum of $f(x)$. 
While it is possible to implement such penalties, it typically does need additional qubits. 
For linear constraints on the string $x$ one should expect to add $O(n)$ additional qubits to the problem definition to implement the penalty, thereby significantly enlarging the search space that the algorithm needs to explore.  
In practice, therefore, the penalty function approach is often not practical, as the disadvantages of the larger search space counterbalance any possible advantages that one gets from rephrasing it as an unconstrained problem. 

\section{Extending QAOA}\label{section:xQAOA}
To deal with the constraints on optimization problems we look at extensions of the QAOA with more general transformations $B, C$ and initial state $\ket{s}$ and we refer to these more general optimization algorithms as \xQAOA. 
In this paper, we focus on \xQAOA that biases the final state towards valid solutions for problems with linear constraints.
Generalizations of QAOA have previously been introduced in \cite{hadfield_etal-2017} where they rename it as \emph{Quantum Alternating Operator Ansatz} Algorithms. 
Previous works \cite{hadfield_etal-2017, wang_etal-2019,bartschi_eidenbenz-2020} have used variants of the mixing Hamiltonian $B$ to preserve the feasible subspace for specific constraint problems, at the cost of additional complexity for initial state preparation, and/or mixer depth. 
In contrast, our algorithm has a constant depth mixer and initial state preparation circuit, just like the original QAOA.

In general, preparing the state, that is the superposition of all feasible states, or just rotating around it, is nontrivial— and requires complex circuits. Rather than constricting our ansatz to the feasible subspace, we attempt to keep it \emph{nearby}.

We introduce two mixers that bias the xQAOA towards  given bit-wise marginal distributions. The \emph{hourglass mixer} biases towards the product distribution of the given marginals. To induce arbitrary (anti-)correlations between any two pairs of bits (while respecting the marginals), we also introduce the \emph{copula mixer}. 
Our implementation for the hourglass mixer is very similar to that in \emph{Warm-Start QAOA} described earlier by Egger, Mareček, Woerner \cite{egger_etal-2020}.

\subsection{Biased Initial States}

In standard QAOA, we start with an initial state that is the uniform superposition over all bitstrings $\ket{+}^{\otimes n} = \frac{1}{\sqrt{2^n}} \sum_{x\in \{0,1\}^n} \ket{x}$. 
Instead, we start from a state that is more biased towards good solutions. 
Consider a classical distribution $p = (p_1, \dots, p_n )$ such that bit $i$ is $1$ with independent probability $p_i$.
We can use this classical distribution to inspire a bias of $p_i$ for qubit $i$ in the following way. Define
\begin{align}
\begin{cases}        
\ket{p_i} :=  \sqrt{1-p_i}\ket{0} + \sqrt{p_i}\ket{1},  \\
    \ket{p_i^\perp} :=  -\sqrt{p_i}\ket{0} + \sqrt{1-p_i}\ket{1}
    \end{cases}
\end{align}
such that $\ket{p_i^\perp}$ is orthogonal to $\ket{p_i}$. 
The entire quantum state for $p$ thus corresponds to the tensor product
\begin{align} \label{eqn:p_state}
    \ket{p} = \ket{p_1}\otimes\ket{p_2}\otimes\cdots \otimes\ket{p_n}. 
\end{align}

\subsection{Single Qubit Mixers \label{sec:hourglass}}
Typical QAOA uses the $\Xgate$-mixer $\sum_i \Xgate_i$. This $\Xgate$-mixer preserves the standard initial state $\ket{+}^{\otimes n} =  \frac{1}{\sqrt{2^n}} \sum_{x=0}^{n-1} \ket{x}$. The hourglass mixer is defined such that it does the same with the biased initial state $\ket{p}$ defined above (\ref{eqn:p_state}).

The \emph{hourglass mixer} is a single qubit gate, that is especially well suited for problems with constraints and/or biases on the possible input states.
While the $\Xgate$ gate has the uniform superposition $(\ket{0}+\ket{1})/\sqrt{2}$, and $(\ket{0}-\ket{1})/\sqrt{2}$ as its eigenstates, and the $\Zgate$ gate has $\ket{0}$ and $\ket{1}$ as its eigenstates, the hourglass mixer is designed to have as its eigenstate ``something in between''. The Hourglass mixer is the same as the mixer used in \cite{egger_etal-2020}.

\paragraph{Single Qubit Hourglass Mixer}
For probability $p_i \in [0,1]$ the hourglass mixer $\Hourglass_{p_i}$ has $\ket{p_i}$ and $\ket{p_i^\perp}$ as its eigenstates, with $-1$ and $+1$ eigenvalues respectively:
\begin{align}
    \Hourglass_{p_i} \ket{p_i}       = -\ket{p_i} \text{~and~}
    \Hourglass_{p_i} \ket{p_i^\perp} = +\ket{p_i}.
\end{align}
The hourglass mixer can be written as a linear combination of the $\Xgate$ and $\Zgate$ gates, hence the ``hourglass'' notation $\Hourglass$, which is an $\Xgate$ and a $\Zgate$ overlayed:
\begin{align}\label{eqn:hourglass}
    \Hourglass_{p_i} & =
    -(1-2p_i)\Zgate - 2\sqrt{p_i(1-p_i)} \Xgate\\
     & = -
    \begin{pmatrix}
        1 - 2p_i    & 2\sqrt{p_i(1-p_i)} \\
        2\sqrt{p_i (1-p_i)} & -(1 - 2p_i)
    \end{pmatrix}
\end{align}
such that $\Hourglass_0 = -\Zgate$ and $\Hourglass_{1/2}=-\Xgate$.
Defining  $\phi_{p_i}=2\sin^{-1}(\sqrt{p_i})$ this transformation can be written as
\begin{align}
    \Hourglass_{p_i} & =
    -\begin{pmatrix}
        \cos \phi_{p_i} & \sin \phi_{p_i}  \\
        \sin \phi_{p_i} & -\cos \phi_{p_i}
    \end{pmatrix} \\
&                      =
    -\begin{pmatrix}
        1 - 2\sin^2(\tfrac{\phi_{p_i}}{2})    & 2\cos(\tfrac{\phi_{p_i}}{2})\sin(\tfrac{\phi_{p_i}}{2}) \\
        2\cos(\tfrac{\phi_{p_i}}{2})\sin(\tfrac{\phi_{p_i}}{2}) & -(1 - 2\sin^2(\tfrac{\phi_{p_i}}{2}))
    \end{pmatrix}.
\end{align}
This gate can easily be implemented in a quantum circuit with the Rotate-Y gate $\Rgate_\Ygate(\phi_{p_i}) = \e^{-\i\phi_{p_i}\Ygate/2}$, its adjoint, and a $\Zgate$ gate in between:
\begin{align}\label{eq:hourglass_hamiltonian}
    \Hourglass_{p_i} = \Rgate_\Ygate(\phi_{p_i})\Zgate \Rgate_\Ygate(\phi_{p_i})^\dagger.
\end{align}
Often, we will be using the $\beta$-wighted Hourglass mixer, $\e^{-\i \beta \Hourglass_{p_i}}$, which can be similarly implemented as
\begin{align}
    \e^{-\i\beta\Hourglass_{p_i}} = \Rgate_\Ygate(\phi_{p_i})\e^{-\i\beta\Zgate} \Rgate_\Ygate(\phi_{p_i})^\dagger.
\end{align}



\paragraph{Final Biased Mixing Hamiltonian}
We design our mixing Hamiltonian $B$ to preserve the biased state $\ket{p}$ as the ground state. We can do this with the Hourglass mixer defined above in Equation~\ref{eqn:hourglass}. 
We simply combine each of the single-qubit Hourglass Hamiltonians for each qubit with its respective probability $p_i$.
\begin{align}
    B_{\Hourglass_p} &= \sum_{i=1}^n \Hourglass_{p_i}^{(i)} \\
    U^{B_{\Hourglass_p}}(\beta) &= \prod_{i=1}^n \e^{-\i\beta\Hourglass_{p_i}^{(i)}},
\end{align}
where the superscript $(i)$ denotes that the operator is acting on the $i$th qubit.
As a result $\ket{p}$ is indeed the unique ground state of $B_{\Hourglass_p}$. 
Note that setting all $p_i = \frac{1}{2}$ implies $\Hourglass_{p_i} = -\Xgate$, which brings us back to standard QAOA.

\subsection{Two Qubit Mixing}
Given a probability distribution $p_{12}:\{0,1\}^2\rightarrow \RR$ for two random variable bits $X_1, X_2$ such that $\Pr[X_1{=}x_1,X_2{=}x_2] = p_{12}(x_1,x_2)$, how can we implement a mixer that respects this distribution? 
We define the following quantum two-qubit state corresponding to the classical distribution
\begin{align}\label{eqn:corr_state}
    \ket{p_{12}} = \sum_{x\in\{0,1\}^2}\sqrt{p_{12}(x_1,x_2)}\ket{x_1x_2}.
\end{align}
Next, we will construct the $\Rgate_{p_{12}}$ gate, which will prepare $\ket{p_{12}}$ from the $\ket{00}$ state:
$\Rgate_{p_{12}} : \ket{00} \mapsto \ket{p_{12}}$.

First, we define the marginal and conditional probabilities
\begin{align}
p_1 & := \Pr[X_1{=}1] = p_{12}(1,0) + p_{12}(1,1), \\
p_{2| 1} &:= \Pr[X_2{=}1| X_1{=}1]= {p_{12}(1,1)}/{p_1},\\
    p_{2| \neg 1} &:= \Pr[X_2{=}1| X_1{=}0] = {p_{12}(0, 1)}/{(1 - p_1)}.
\end{align}
Next, we define the corresponding angles
\begin{align}
\phi_{p_1} & = 2\sin^{-1}(\sqrt{p_1}),\\
\phi_{p_{2|1}} & = 2\sin^{-1}(\sqrt{p_{2| 1}}),\\
\phi_{p_{2|\neg 1}} & = 2\sin^{-1}(\sqrt{p_{2|\neg 1}}).
\end{align}
The circuit for $\Rgate_{p_{12}}$ uses these angles as shown below
\begin{align}\label{eq:corr_rotation}
\begin{array}{c}
\Qcircuit @C=1em @R=1em {
     &
        \gate{R_\Ygate(\phi_{p_2})} &
            \ctrl{1} & 
                \ctrlo{1} &
                    \qw \\
    & 
        \qw &
            \gate{R_\Ygate(\phi_{p_{2|1}})} &
                \gate{R_\Ygate(\phi_{p_{2|\neg 1}})} &
                    \qw
}
\end{array}
\end{align}


We use the $\Rgate_{p_{12}}$ gate to construct our 2-qubit mixing Hamiltonian. Just like in the Hourglass mixer construction we design our mixing Hamiltonian $B_{p_{12}}$ so that $\ket{p_{12}}$ is the ground state
\begin{align}\label{eq:corr_mixer_properties}
    B_{p_{12}} \ket{p_{12}} = -2\ket{p_{12}}, \quad \bra{p_{12}^\perp} B_{p_{12}} \ket{p_{12}^\perp} \ge 0
\end{align}
where $\ket{p_{12}^\perp}$ denotes any state orthogonal to $\ket{p_{12}}$.

We can accomplish Eq.~\ref{eq:corr_mixer_properties} by combining the rotation gate $\Rgate_{p_{12}}$ 
(Eq.~\ref{eq:corr_rotation}) to change bases, and a diagonal operator $D$ to assign eigenvalues, i.e.\
$B_{p_{12}} = \Rgate_{p_{12}} D \Rgate_{p_{12}}^\dagger$
with 
\begin{align}
D & = -\Zgate_1 -\Zgate_2  = \Bigg( \begin{smallmatrix}
        -2 \\  & 0 \\  & & 0 \\ & & & 2
    \end{smallmatrix}\Bigg).
\end{align}
As a result, our parametrized $2$ qubit mixing unitary is implemented as
\begin{align}\label{eq:corr_mixer}
    \e^{-\i\beta B_{p_{12}}} &= \Rgate_{p_{12}} \e^{i\beta Z_1} \e^{i\beta Z_2} \Rgate_{p_{12}}^\dagger
\end{align}
\begin{align}
&=
\begin{array}{c}
\Qcircuit @C=1em @R=1em {
    & \multigate{1}{\Rgate_{p_{12}}^\dagger} & \gate{R_\Zgate(2\beta)} & \multigate{1}{\Rgate_{p_{12}}} & \qw\\
    & \ghost{\Rgate_{p_{12}}^\dagger} & \gate{R_\Zgate(2\beta)} & \ghost{\Rgate_{p_{12}}} &\qw
}
\end{array}
\end{align}
Note that when the random variables $X_1, X_2$ are independent, $R_{p_{12}}$ is equivalent to $\Rgate_\Ygate(\phi_{p_1}) \otimes \Rgate_\Ygate(\phi_{p_2})$, and this mixer is equivalent to the Hourglass mixer in  \S\ref{sec:hourglass}, Equation~\ref{eq:hourglass_hamiltonian} as then $B_{p_{12}} = \Hourglass_{p_1} + \Hourglass_{p_2}$. 

\subsection{Two Bit Copulas}
Next we explain how for two marginal distributions for two bits we can use a parameterized joint distribution, a \emph{bivariate copula} $c_\theta$, to define a $2$-qubit mixer that favors the qubits to be correlated ($\theta \in (0,1]$) or anti-correlated ($\theta\in[-1,0)$). 

Assume two random bit variables $X_1$ and $X_2$ with marginal probabilities $\Pr[X_1{=}1] = p_1$, $\Pr[X_2{=}1] = p_2$ and hence with the expectations $\E[X_1] = p_1$ and $\E[X_2]=p_2$. 
A copula $c_\theta:\{0,1\}^2\rightarrow \RR$ is a probability distribution $c_\theta(x_1,x_2) = \Pr[X_1{=}x_1, X_2{=}x_2]$ such that the marginals again obey:
\begin{align}
c_\theta(1,0) + c_\theta(1,1) & = p_1, \\
c_\theta(0,1) + c_\theta(1,1) & = p_2.
\end{align}
For copulas $c_\theta$ with covariance $\E[X_1X_2]-\E[X_1]\E[X_2] = \Delta$ it must hold that 
\begin{align}\label{eq:copula_constraints}
\begin{cases}
c_\theta(0,0) & = (1-p_1)(1-p_2) + \Delta  \\ 
c_\theta(0,1) & = (1-p_1)p_2 - \Delta\\
c_\theta(1,0) & = p_1(1-p_2) - \Delta \\
c_\theta(1,1) & = p_1p_2 + \Delta.  
\end{cases}
\end{align}
Note that for $\Delta=0$ we have independent $X_1$ and $X_2$, which should coincide with the $\theta{=}0$ copula with probabilities $c_0(1,1) = p_1p_2$, et cetera.  
For $\Delta$ to define a valid distribution it is required that
\begin{align}
\max\{0, p_1+p_2-1\} & \leq   p_1p_2+\Delta  \leq \min\{p_1,p_2\}. \label{eq:valid_copula}
\end{align}
To satisfy this requirement for all cases we define the following covariance, which is inspired by the Farlie-Gumbel-Morgenstern (FGM) copula \cite{F,G,M}:
\begin{align}\label{eq:FGM}
\Delta  & = \theta p_1p_2(1-p_1)(1-p_2).
\end{align}
It is straightforward to verify that for all $\theta\in[-1,+1]$ this definition satisfies the constraints of Equation~\ref{eq:valid_copula}.

\subsection{Two Qubit Copula Mixers} \label{sec:copula_mixers}
Combining the above two sections, given the marginal distributions for two bits $X_1, X_2$, we can induce an arbitrary (anti-)correlation $\theta$, while still maintaining the marginals. 
Given marginals $p_1 = \Pr[X_1{=}1]$ and $p_2 = \Pr[X_2{=}1]$, we can fully define the $2$-qubit joint distribution by choosing a value for the correlation parameter $\theta \in [-1, 1]$ in combination with the Equalities~\ref{eq:copula_constraints} and \ref{eq:FGM}.
Because the two qubit gate $\Rgate_{p_{12}}$ is determined by this joint distribution, we will now denote it as
$\Rgate(p_1, p_2, \theta)$, or $\Rgate^{(1,2)}(p_1, p_2, \theta)$ to indicate that it is acting on the $1$st and $2$nd qubit.

Specifically, we can calculate the values for $p_1, p_{2|1}, p_{2|\neg 1}$ as follows:
\begin{align}
p_1 & = \Pr[X_1{=}1], \\
p_{2| 1} &= \Pr[X_2{=}1| X_1{=}1]= p_2 + \theta p_2(1-p_1)(1-p_2),\\
    p_{2| \neg 1} &= \Pr[X_2{=}1| X_1{=}0] = p_2- \theta p_1p_2(1-p_2).
\end{align}
These probabilities are used to then construct the circuit for $\Rgate_{p_{12}}$ as shown in Equation~\ref{eq:corr_rotation}.

We define our Copula mixing Hamiltonian parameterized by $(p_1, p_2, \theta)$ as
\begin{align}
    \Cop_{12} & = \Cop(p_1, p_2, \theta)\\ & = -\Rgate(p_1, p_2, \theta) (\Zgate^{(1)} + \Zgate^{(2)}) \Rgate(p_1, p_2, \theta)
\end{align}
with unitary evolution
\begin{align}
U^{\Cop_{12}} & =  \e^{-\i\beta\Cop(p_1, p_2, \theta)}\\
    &= \Rgate(p_1, p_2, \theta) \Rgate^{(1)}_\Zgate(-2\beta) \Rgate_\Zgate^{(2)}(-2\beta) \Rgate(p_1, p_2, \theta)^\dagger.
\end{align}


\subsection{Many Qubit Copula Mixer} 
To induce (anti-)correlations between pairs of qubits, we can sum up the different $2$-qubit Copula Hamiltonians for each relevant pair. 
Let ${p} = (p_1, \dots, p_n)$, and $\Theta = \{(a_i, b_i, \theta_i)\}_i$ be a set of 3-tuples indicating to correlate bits $(a_i, b_i)$ with correlation parameter $\theta_i$. 
The $n$-qubit Copula mixing Hamiltonian can then be written as
\begin{align}\label{eq:multi_copula_ham}
    \Cop_n({p}, {\Theta}) = \sum_{(a, b, \theta) \in \Theta}
    \Cop^{(a,b)}({p}_a, {p}_b, \theta)
\end{align}
Exactly implementing the unitary  operation $\e^{-i\beta \Cop_n}$ with 2-qubit gates is complicated though by the fact that the terms in the Hamiltonian typically do not commute. 
Only in the case that $\Theta$ consists of disjoint pairs of qubits can we implement each term in parallel, as we did for the Hourglass mixer. 

One can approximate the unitary evolution of Hamiltonian~(\ref{eq:multi_copula_ham}) via Trotterization \cite{trotterization}, or a first order approximation thereof that has a similar functionality, as explored for the mixers proposed in \cite{hadfield_etal-2017}. 
In general, the Trotterization method consists of partitioning the terms of the Hamiltonian into those that commute and hence are easier to implement in series.
We will make use of this method in the next section.

\begin{figure}
    \centering
    \[
    \begin{array}{c}
    \Qcircuit @C=.5em @R=.5em {
        & \ustick{U^{\CopOn}} & & &   \ustick{U^{\CopEn}}    \\
        & \multigate{1}{U^{\Cop_{12}}} & \qw
            & \qw
                & \qw & \sgate{U^{\Cop_{6 1}}}{5} & \qw
        \\
        & \ghost{U^{\Cop_{12}}} & \qw 
            & \qw & \multigate{1}{U^{\Cop_{23}}} & \qw & \qw
        \\
        & \multigate{1}{U^{\Cop_{3 4}}} & \qw 
            & \qw & \ghost{U^{\Cop_{23}}}
            & \qw & \qw
        \\
        & \ghost{U^{\Cop_{3 4}}} & \qw 
            & \qw & \multigate{1}{U^{\Cop_{45}}} &  \qw& \qw
        \\
        & \multigate{1}{U^{\Cop_{56}}} & \qw
            & \qw & \ghost{U^{\Cop_{45}}} & \qw & \qw\\
        & \ghost{U^{\Cop_{56}}} & \qw 
            & \qw
                & \qw &  \gate{U^{\Cop_{6 1}}} &  \qw
        \gategroup{2}{2}{7}{2}{.7em}{^\}}
        \gategroup{2}{4}{7}{6}{.7em}{^\}}
    } 
    \end{array}
    \]
    \caption{ Quantum circuit implementation for the \emph{partitioned Copula ring} mixer $\tilde{U}^{\Copn}$ of Equation~\ref{eq:part_ring_copula_unitary} for $6$ qubits. We denote $U^{\Cop_{ab}} = U^{\Cop(p_a, p_b, \theta_a)}$.}
    \label{fig:ring_copula}
\end{figure}
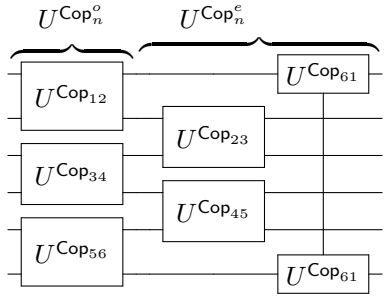

\paragraph{Ring Copula $n$-qubit mixer.}\label{sec:ring_copula}
Consider the scenario when we would like to enforce some (anti-)correlation between all pairs of neighboring qubits in a ``ring''
\begin{align}
    \Copn({p}, {\theta}) = \sum_{i \in \{1,\dots,n\}} \Cop^{(i, i{+1})}({p}_i, {p}_{i+1}, {\theta}_i)
\end{align}
for ${\theta} \in [-1, 1]^n$.
For simplicity, assume $n$ is even. 
Implementing the unitary evolution $U^{\Copn}(\beta) = \e^{-\i\beta \Copn}$ with 2-qubit gates is nontrivial, since the terms do not commute for nonzero $\theta_i$. 
However, we can partition the Hamiltonian into two sets, ``odd'' and ``even'', such that all elements within a set commute:
\begin{align}
    \Copn({p}, {\theta}) =  \CopEn({p}, {\theta}) +\CopOn({p}, {\theta})
\end{align}
where
\begin{align}
    \CopOn({p}, {\theta}) &=
        \sum_{i \in \{1,3,\dots,n-1\}} \Cop^{(i, i{+1})}({p}_i, {p}_{i+1}, {\theta}_i)\\
        \CopEn({p}, {\theta}) &= 
        \sum_{i \in \{2,4,\dots,n\}} \Cop^{(i, i{+1})}({p}_i, {p}_{i+1}, {\theta}_i).
\end{align}
and used a convention where $\Cop^{(n,n+1)}({p}_n,{p}_{n+1},\theta_n) = 
\Cop^{(n,1)}({p}_n,{p}_{1},\theta_n)$. 
With this, we can implement the approximating \emph{partitioned Copula ring} mixer
\begin{align}\label{eq:part_ring_copula_unitary}
    \tilde{U}^{\Copn}(\beta) :=
(\e^{-\i\beta\CopEn({p}, {\theta})})(\e^{-\i\beta\CopOn({p}, {\theta})}).
\end{align}
Each of these two operators can be implemented in constant depth since their terms commute, for example:
\begin{align}
    \e^{-\i\beta\CopOn({p}, {\theta})} &= 
        \prod_{i\in\{1,3,\dots,n-1\}} \e^{-\i\beta\Cop^{(i, i{+1})}({p}_i, {p}_{i+1}, {\theta}_i)}.
\end{align}
See Figure~\ref{fig:ring_copula} for the circuit implementation of $\tilde{U}^{\Copn}$.

%% file: Knapsack_Background.tex
\section{Knapsack Problem Background}\label{section:KP}
\subsection{Knapsack Problem Definition and Hardness}
\label{sec:kp-definition}
We remind the reader that instances $(n,v,w,c)$ of the Knapsack Problem (KP) of size $n$ are defined as follows. 
Given $n$ items of weights $w_1, \dots, w_n \in \NN$, values $v_1, \dots, v_n \in \NN$, and a capacity $c \in \NN$, find a subset of items that maximizes the combined value while keeping the sum of their weights under capacity. 
That is, find a binary string
$x^* = (x_1, \dots, x_n) \in \{0,1\}^n$ such that 
\begin{align}\label{eq:KPx}
x^* & = \arg\max_{x\in\{0,1\}^n} (x_1v_1+\cdots + x_nv_n)
\end{align}
under the constraint
\begin{align}\label{eq:KPw}
x_1w_1+\cdots + x_nw_n \leq c.
\end{align} 


%

The KP has many connections with real-world problems. In fact, we were initially motivated by a cybersecurity problem that naturally extends to a KP \cite{eldefrawy_etal-2007}. 
However, phrased as a decision problem, the KP is well known to be NP-complete, hence we should not expect to find an efficient perfect quantum algorithm for solving KP. 
Instead, we will focus on heuristic optimization algorithms that perform well on certain distributions of instances.
Since we are comparing against QAOA, we
consider an approximation algorithm for KP \emph{efficient} if it runs in time and space linear or almost linear in $n$.

\subsection{Classical Heuristics for KP} \label{sec:classical_algorithms}
\begin{figure}
\begin{minipage}[T]{.5\textwidth}
\centering
\begin{algorithm}[H]
\label{alg:LG}
\SetAlgoLined
\textbf{Input} : $n\in \ZZ^+,\  v, w\in\RR^n,\ c\in \RR^+$ \\
\textbf{Output} : $x \in \{0, 1\}^n$\\
    $r_i \gets v_i/w_i$ for all $i\in\{1,\dots,n\}$\\
    Sort the indices in a non-increasing order by the 
    $r_i$ values. Ties in $r_i$ go to the smaller $i$. \\
    Let $\sigma$ denote this permutation, so $r_{\sigma^{-1}(1)} \geq r_{\sigma^{-1}(2)} \geq \dots ,\geq r_{\sigma^{-1}(n)}$.\\
    $c' \gets c - w_{\sigma^{-1}(1)}$\\
    $j \gets 1$\\
    $x \gets (0, 0, \dots, 0) \in \{0,1\}^n$\\
    \While{$c'>0$ \text{and} $j\leq n$}{
      $x_{\sigma^{-1}(j)} \gets 1$\\
         $j \gets j+1$\\
        $c' \gets c' - w_{\sigma^{-1}(j)}$
    }
    \Return $x$.    
\caption{ \textsc{Lazy Greedy}$(n, v, w, c)$}
\end{algorithm}
\end{minipage}
\begin{minipage}[T]{.5\textwidth}
\centering
\begin{algorithm}[H]
\label{alg:VG}
\SetAlgoLined
\textbf{Input} : $n\in \ZZ^+,\  v, w\in\RR^n,\ c\in \RR^+$ \\
\textbf{Output} : $x \in \{0, 1\}^n$\\
    $r_i \gets v_i/w_i$ for all $i\in\{1,\dots,n\}$\\
    Sort the indices in a non-increasing order by the 
    $r_i$ values. Ties in $r_i$ go to the smaller $i$. \\
    Let $\sigma$ denote this permutation, so $r_{\sigma^{-1}(1)} \geq r_{\sigma^{-1}(2)} \geq \dots ,\geq r_{\sigma^{-1}(n)}$.\\
    $c' \gets c$\\
    $j \gets 1$\\
    $x \gets (0, 0, \dots, 0) \in \{0,1\}^n$\\
    \While{$c'>0$ \text{and} $j\leq n$}{
      \While{$c' < w_{\sigma^{-1}(j)}$}{$j \gets j+1$\\}
      $x_{\sigma^{-1}(j)} \gets 1$\\
         $j \gets j+1$\\
        $c' \gets c' - w_{\sigma^{-1}(j)}$
    }
    \Return $x$.    
\caption{ \textsc{Very Greedy}$(n, v, w, c)$}
\end{algorithm}
\end{minipage}
    \caption{Both greedy algorithms are described above.
    The only difference is that the \textsc{VeryGreedy} algorithm
    runs through the entire sorted list of items.}
    \label{fig:GreedyAlgs}
\end{figure}
Here we describe the classical algorithms we used as a baseline benchmark to compare the performance of our quantum algorithms against. We implemented every classical algorithm discussed below in Python 3.

\paragraph{Lazy Greedy Algorithm (LG)}
For each item $i$ define the value per unit weight by the ratio
$r_i = \frac{v_i}{w_i}$.
	The \emph{Lazy Greedy} algorithm sorts the inputs by their ratio $r_i$ in non-increasing order. 
	Next, it picks items until the total weight is under capacity $c$ and the next item in the sorted list would push the total weight to be over capacity.
	Though there are potentially
	other items down the list which could fit (smaller, less efficient ones), the algorithm simply stops and ends.
	If we sort the binary vector representing
	the items by decreasing efficiency, the
	returned vector is a contiguous chain of $1$'s:
	$(1, 1, \dots, 1, 0, \dots, 0) \in \{0,1\}^n$. 

\paragraph{Very Greedy Algorithm (VG)}
	The \emph{Very Greedy} algorithm starts in the same way as the above Lazy Greedy algorithm, but it looks at all items until no	more items can be added.
	In other words, it sorts the items by their efficiency.
	Then, it picks the most efficient remaining items while staying
	under the capacity.
	Assuming the input is sorted by efficiency, this algorithm is linear
	time and space in the number of items and bit length ($n$).
	Hence, it is feasible for a 
	large problem space but will rarely pick an optimal solution.

\begin{figure}
\begin{minipage}[T]{.5\textwidth}
\centering
\begin{algorithm}[H]
\label{alg:SA}
\SetAlgoLined
\textbf{Input} : $n, s\in \ZZ^+,\  v, w\in\RR^n,\ c, T\in \RR^+$ \\
\textbf{Output} : $x \in \{0, 1\}^n$\\
    $x \gets \textsc{LazyGreedy}(n, v, w, c)$.\\
    $\text{value} \gets x_1v_1 + \dots + x_nv_n$.\\
    $x_{max} \gets x$.\\
    \For{$i \gets \{0, \dots, s-1\}$}{
        \Repeat{$y_1w_1 + \dots + y_nw_n \leq c$}{Sample $j$ uniformly at random from $\{1, \dots, n\}$.\\
        $y \gets x \oplus \mathds{1}_j$. \\}
        $\Delta \gets y_1v_1 + \dots + y_nv_n - \text{value}$.\\
        \If{$\Delta > 0$}{$x \gets y$.\\
        $x_{max} \gets x$.\\
        Set $\text{value} \gets y_1v_1 + \dots + y_nv_n$.}
        \Else{$x \gets y$ with probability $e^{\Delta/T}$.\\
        $x_{max} \gets x$.\\
        Set $\text{value} \gets x_1v_1 + \dots + x_nv_n$.}
    }
    \Return $x_{max}$.    
\caption{ \textsc{Simulated Annealing}$(n, v, w, c)$}
\end{algorithm}
\end{minipage}
\begin{minipage}[T]{.5\textwidth}	
\centering
\begin{algorithm}[H]
\label{alg:GSA}
\SetAlgoLined
\textbf{Input} : $n, s\in \ZZ^+,\  v, w\in\RR^n,\ c, T\in \RR^+$ \\
\textbf{Output} : $x \in \{0, 1\}^n$\\
    $x \gets \textsc{Lazy Greedy}(n, v, w, c)$.\\
    $\text{value} \gets x_1v_1 + \dots + x_nv_n$.\\
    $x_{max} \gets x$.\\
    \For{$i \gets \{0, \dots, s-1\}$}{
        \Repeat{$y_1w_1 + \dots + y_nw_n \leq c$}{Sample $j$ uniformly at random from $\{1, \dots, n\}$.\\
        $y \gets x \oplus \text{Bern}(1/n)^n$. \\}
        $\Delta \gets y_1v_1 + \dots + y_nv_n - \text{value}$.\\
        \If{$\Delta > 0$}{$x \gets y$.\\
        $x_{max} \gets x$.\\
        Set $\text{value} \gets y_1v_1 + \dots + y_nv_n$.}
        \Else{$x \gets y$ with probability $e^{\Delta/T}$.\\
        $x_{max} \gets x$.\\
        Set $\text{value} \gets x_1v_1 + \dots + x_nv_n$.}
    }
    \Return $x_{max}$.    
\caption{ \textsc{Global Simulated Annealing}$(n, v, w, c)$}
\end{algorithm}
\end{minipage}
    \caption{Simulated annealing and its global version are described in
    Algorithms~\ref{alg:SA} and ~\ref{alg:GSA}, respectively.
    The only difference between the two is how the two algorithms sample nodes
    during their respective random walks. Simulated annealing
    samples a valid neighbor uniformly at random while global SA
    samples a node by the Bernoulli product distribution, which we denote as $\text{Bern}(1/n)^n$. (The Bernoulli product distribution $\text{Bern}(1/n)^n$ flips each bit with probability $1/n$.)
    The latter takes one step in expectation.}
    \label{fig:SAalgs}
\end{figure}
\paragraph{Simulated Annealing}
Simulated annealing is a random walk over all
		valid collections of items, i.e., under capacity,
		which iterates the following step:
		Say the walk is at node $x \in \{0,1\}^n$ at time $i$. The algorithm
		picks a uniformly random, valid (under capacity) neighboring node.
		It moves to this neighbor if the value increases.
		Otherwise, it moves to the neighbor with probability $e^{\Delta/T}$ where
		$\Delta < 0$ is the decrease in value and $T>0$ is a fixed temperature for all steps.
		
\paragraph{Global Simulated Annealing (GSA)}
 This algorithm is a different version
 		of simulated annealing. In the middle of the random walk, regular simulated
 		annealing looks at each of its neighbors and picks one uniformly at random.
 		This global version of simulated annealing, however, can move to any point in
 		the graph. If we represent a node as a binary string $x \in \{0,1\}^n$,
 		the algorithm's move step is to flip bit $x_i$ with probability $1/n$ for
 		each $i$. Therefore, the algorithm flips one bit in expectation, but can move to
 		see any node in a given move step. Once a node is sampled, the algorithm moves
 		to it if the value increases.
		Otherwise, it moves to the node with probability $e^{\Delta/T}$ where
		$\Delta < 0$ is the decrease in value and $T>0$ is a fixed temperature.
		Note that we allow GSA to move to
		invalid nodes. When this occurs, we
		set that node's score to $0$.
		We do this for a fair comparison
		with our QAOA algorithms.

\subsection{Less Practical Algorithms for KP}

For completeness, we briefly describe other less practical  classical algorithms 
which solve KP. Less practical in this context means that such algorithms require non-linear, and sometimes even exponential running time.

\paragraph{Brute Force}
		The brute force algorithm (i.e., exhaustive search) searches all possible $2^n$ combinations
		and outputs the highest score, Equation~\ref{eq:KPx}, which is below capacity,
		Equation~\ref{eq:KPw}. This algorithm is clearly impractical 
		because it requires exponential time in the number of items.

\paragraph{Dynamic Programming}
		There are multiple ways to solve KP instances exactly with
		dynamic programming (DP). The two main methods result in
		either $n^2 \cdot \max_i (v_i)$ time and space or
		$n\cdot c$ time and space ($c$ being the capacity).
		For efficiency reasons, we chose
		the latter and refer to it as \emph{dynamic programming}
		throughout the remainder.
		DP makes a table of size $n\cdot c$
		and finds the \emph{optimal} solution by solving the problem recursively.
		This algorithm is infeasible for large capacities (as we expect to see in
		real-world problems).
		
\paragraph{Fully Polynomial Time Approximation Scheme (FPTAS)}
		This is an approximation algorithm which runs in time and space
		$n^2 \left\lfloor \frac{n}{\varepsilon} \right\rfloor$ and is
		guaranteed to a return a solution which is a
		$(1-\varepsilon)$-fraction of the optimal solution.
		We did not run this algorithm due to its cubic run-time and memory.
		

%% file: QAOA_for_KP.tex
\section{xQAOA for Knapsack Problems}\label{section:QAOAforKP}

In this section, we will first discuss the  limitations of the standard QAOA for solving KP. 
Next, we extend QAOA to a more suitable form that uses the techniques that were described in Section~\ref{section:xQAOA}.

\subsection{Limitations of Standard QAOA for Solving KP}
While standard QAOA often works well on unconstrained optimization problems, it typically performs significantly less well on constrained problems where large parts of the hypercube $\{0,1\}^n$ should be ignored by the algorithm. 
To overcome this problem, one can consider generalizations of QAOA that allow non-standard variants of the Mixing Hamiltonian $B$, the Energy function $C$, or the initial state $\ket{s}$. 
Going forward we will use the notation that for given $B$, $C$, $s$, and $2p$ angles $(\beta_1,\dots,\beta_p,\gamma_1,\dots,\gamma_p)$, the final state is defined and denoted by
\begin{align}
    \ket{\beta,\gamma} & :=
    U^{B}(\beta_p)
    U^{C}(\gamma_p)
    \cdots
    U^{B}(\beta_1)
    U^{C}(\gamma_1)
    \ket{s}.
\end{align}

We remind the reader that a Knapsack instance is defined by problem size $ n\in\NN$, values $v\in\NN^n$, weights $w\in\NN^n$, and capacity threshold $c\in \NN$. 
The goal is to find the optimal string $x$
\begin{align} \label{eq:KP_objective}
x^* & = \arg\max_{x\in\{0,1\}^n} v \cdot x \\
    & \text{subject to }  w \cdot x \leq c.
\end{align}
Equation~\ref{eq:KP_objective} can be written for a quantum state, as minimizing the Hamiltonian
\begin{align}
    C = \sum_{i=1}^n v_i\cdot \Zgate_i.
\end{align}
The corresponding unitary can be implemented in depth $1$:
\begin{align}
    U^C(\gamma) &= U^{v_1}(\gamma)\otimes \cdots \otimes U^{v_n}(\gamma),
    \end{align} 
    where 
    $U^{v_i}(\gamma)\ket{x_i} = \e^{-\i \gamma v_i x_i}\ket{x_i}$.
    
\subsection{Designing xQAOA for Knapsack Problems}
In this section, we will extend QAOA to make it more suitable for Knapsack Problems. 
\subsubsection{Initial State Bias}\label{sec:initial_state_bias}
We propose to use a biased initial state $\ket{p_1,\dots,p_n}$ as opposed to the standard uniform superposition. 
Since all values $v_i$ are positive, we should suspect that the optimal solution has $w\cdot x$ close to the threshold value $c$. 
We thus may want to bias our state so that in expectation we have $w \cdot x = c$. 
\paragraph{Constant Bias State}
Below we define a distribution with a constant bias such that in expectation we reach the constraint threshold.
\begin{align}
p_i & = \frac{c}{\sum_i w_i},\\
 \mathbb{E}[w \cdot x] & = \sum_{i=1}^n w_ip_i = \frac{\sum_i w_i c}{\sum_i w_i} = c. 
 \end{align}
 Note that when $\sum_i w_i < c$ the optimization problem has the trivial solution $x=1^n$, hence we do not have to worry about $p_i$ being ill-defined. 
 
\paragraph{Lazy Greedy Bias}
The \emph{Lazy Greedy algorithm,} as described in \S\ref{sec:classical_algorithms}, sets bits to $1$ in the descending order of $r_i = v_i/w_i$ until flipping a bit to $1$ would result in $w \cdot x$ surpassing the capacity $c$. 
We can describe this deterministic distribution as
\begin{align}p_i =
    \begin{cases}
        1 & \text{if } r_i > r_\text{stop}    \\
        0 & \text{if } r_i \leq r_\text{stop}
    \end{cases}
\end{align}
where $r_\text{stop}$ is the largest $r_i$ such that bit $i$ is set to 0 by LG.

\paragraph{Smoothened Lazy Greedy}
The \emph{Constant Bias} and \emph{Lazy Greedy} distributions both provide instance-specific bias, but at two different extremes. 
The constant bias, treats each bit equally, regardless of the values of $v_i$ or $r_i$. 
On the other hand, Lazy Greedy puts so much bias on each bit that it leads to deterministic distribution. 
Next, we define an initial state that smooths out the Lazy Greedy step function to combine these two distributions.
\begin{figure}
    \centering
    \includegraphics[width=.7\columnwidth]{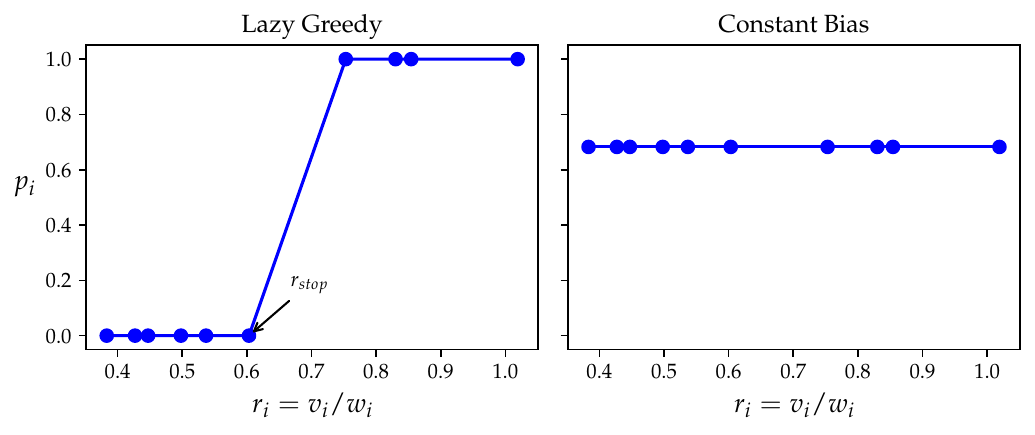}
    \caption{Lazy Greedy distribution and Constaint Bias distibution as a function of ratio $r_i = v_i/w_i$ for
    the Knapsack problem instance: $v{=}(3627,  580, 1835,  246,  364,  674,  840, 1391,  250,  193)$, $w{=}(6012, 1297, 2148,  642,  678,  895, 1012, 1365,  502,  452)$, $c{=}10240$. This problem instance is a scaled version of the \emph{Code Red} instance in \cite{eldefrawy_etal-2007}.
    }
    \label{fig:vgreedy_vs_constant}
\end{figure}
Using the logistic function and free parameters $k$, $C$, $r^*$ we define
\begin{align} \label{eq:logistic}
p_i = \frac{1}{1 + Ce^{-k(r_i - r^*)}}
\end{align}
The following three properties are easy to verify:
\begin{enumerate}
    \item This is a smooth curve in $r_i$ with an inflection  at $r^*$.
    \item In the limit $k \rightarrow \infty$, we get back the Lazy Greedy  distribution if $r^* = r_\text{stop}$.
    \item In the limit $k\rightarrow 0$, we get back the uniform distribution $p_i = \frac{c}{\sum_i w_i}$ if we set 
    $C = {\sum_i w_i}/{c} - 1$.
\end{enumerate}
\begin{figure}[t]
    \centering
    \includegraphics[width=.9\columnwidth]{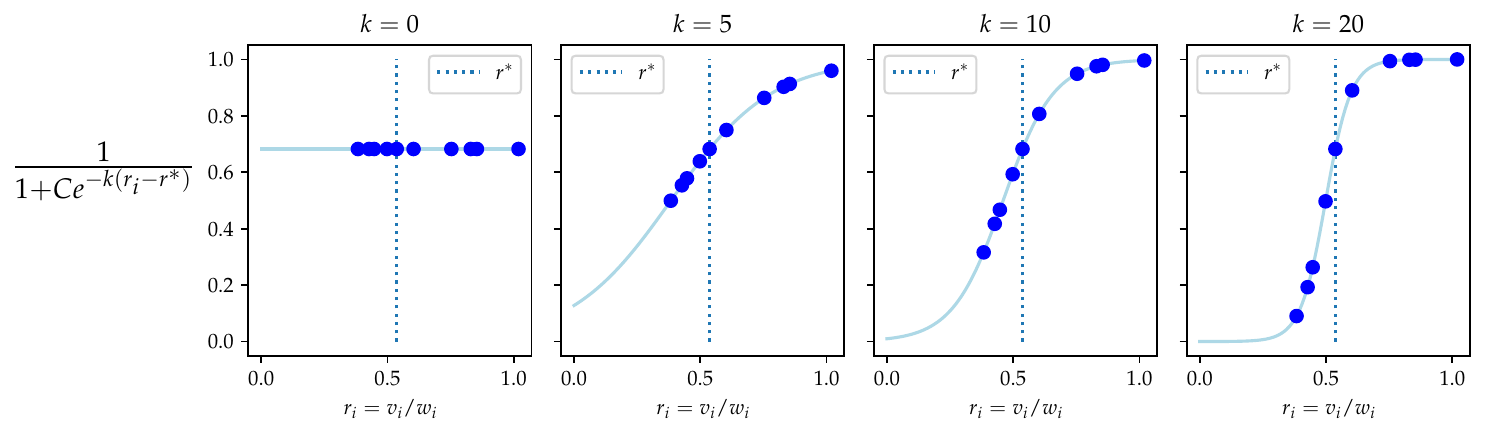}
    \caption{Smoothed out Lazy Greedy distribution using the Logistic function $ p_i = (1 + Ce^{-k(r_i - r^*)})^{-1}$ with $C=\sum_i w_i/c-1$ for different value of $k$. 
    Each blue dot corresponds to a single bit's distribution in the scaled \emph{Code Red} Knapsack Problem instance \cite{eldefrawy_etal-2007}. See the caption of Figure  \ref{fig:vgreedy_vs_constant} for a full description of the problem instance.
    }
    \label{fig:logistic}
\end{figure}
Figure~\ref{fig:logistic} shows how this Logistic distribution changes as we vary $k$ in Equation~\ref{eq:logistic}, which determines the steepness of the function at its inflection point.


\subsubsection{Biased Single Qubit Mixers}
Next, we propose our single qubit mixers using the same biased distribution $\{p_i\}_i$ used for the initial state of Equation~\ref{eq:logistic}. 
By doing so we force the quantum algorithm to stay relatively close to the Lazy Greedy solution, thus avoiding considering unfeasible solutions. 
Specifically, we use the Hourglass mixer of Sec.~\ref{sec:hourglass} to implicitly use a Hamiltonian that has $\ket{p}$ as its ground state.

It should be stressed that because of the absence of qubit-qubit interactions in both the Knapsack cost function as well as the hourglass mixer, the  algorithm thus far does not qualify as a true quantum algorithm, as it only involves single-qubit gates. 
The algorithm can therefore be simulated efficiently classically and should be considered a \emph{quantum-inspired} algorithm. 
Yet we will see in the next section that despite this simplicity it will often outperform standard classical algorithms. 

\subsubsection{Biased Two Qubit Mixers}
The Copula Mixer of Section~\ref{sec:copula_mixers} allows us to extend the Hourglass mixer to induce entanglement in the output state. 

Glover in \cite{Glover} pointed out that rather than evaluating each knapsack item individually just by its ratio $r_i{=}v_i/w_i$, it is sometimes more advantageous to consider jointly multiple items at a time. 
Adding the knapsack items with the very highest ratios may be a good greedy strategy for the first items, however, as the knapsack gets filled up, we run into problems. 
Consider the following example from \cite{Glover} where we want to maximize $2 x_a + 100 x_b$, subject to: $1x_a +  51x_b \leq 51$.
Only one of these two items can fit in the knapsack, however, by using the ratio as the sole heuristic, we will always favor the worse item $x_a$.

Motivated by this example, we propose to anti-correlate bits with ratios that are close.
 Consider the variables $x_1, x_2, \hdots, x_n$ ordered in descending order by ``bang-for-buck" ratio such that $r_i \ge r_{i+1}$. We use the Ring Copula mixer (\S \ref{sec:copula_mixers}) to anti-correlate all pairs of neighboring qubits, by setting the correlation parameter $\theta$ to a negative value for all such pairs.  

 Even though we use the same correlation parameter $\theta$ for each pair, the effect of this correlation will most affect those pairs near the inflection point of the logistic function $r^*$ (Fig.~\ref{fig:logistic}). This is because when marginal distributions of two bits are both already very close to either $0$ or $1$, the copula has very little ``wiggle room'' which can be interpreted as the range of $\Delta$ that satisfy Eq.~\ref{eq:valid_copula}.
As will be shown in our experimental results below, maximally anti-correlating neighboring bits ($\theta{=}{-}1$) with the copula mixer typically outperformed the standard hourglass mixer.

%% file: Experiments_and_Results.tex
\section{Experiments and Results}\label{section:results}
\subsection{Hard KP Instances}\label{ssec:instances}

We relied on analysis in previous work \cite{pisinger_05} to generate random KP instances. 
Specifically, we chose five distributions from Pisinger's work investigating distributions that generate hard Knapsack  instances.
We chose these distributions because they are documented as being hard for state-of-the-art families of dynamic programming algorithms and they have clear, interesting challenges for greedy algorithms.

In all of the following distributions, we first sampled $10$ random weights and values: $(v_1, \dots, v_{10})$ and
$(w_1, \dots, w_{10})$. Then, we set the capacity as
$c \gets \left\lceil\alpha\cdot\left(\sum_i w_i\right)/100 \right\rceil$ where $\alpha$ is
a uniformly random value in $\{25, 26, \dots, 75\}$. Finally, the knapsack instance is to maximize $x\cdot v$ subject to $x\cdot w \leq c$ and $x \in \{0,1\}^n$.

\subsubsection{Strongly Correlated Distribution} The weights $w_j$ of the ``Strong'' distribution are sampled randomly from $\{1, 2, \dots ,1000\}$ and its values are set to $v_j = w_j + 1000$. Such instances correspond to a real-life situation where the return is proportional to the investment plus
some fixed charge for each project. The strongly correlated instances are hard to solve for two reasons:
(a) The instances are ill-conditioned in the sense that there is a large gap between the continuous and integer solution of the problem. (b) Sorting the items according to decreasing efficiencies corresponds to sorting according to the weights. Thus for any small interval of the ordered items (i.e., a ``core''), there is a limited variation in the weights, making it difficult to satisfy the capacity constraint with equality.

\subsubsection{Inversely Strongly Correlated Distribution}
The ``Inv.\ Strong'' distribution picks the values, $v_i$, randomly from $\{1,2, \dots, 1000\}$ then the
weights randomly from $\{v_j + 98, v_j + 102\}$. 

\subsubsection{Profit Distribution}
``Profit,'' or ``profit ceiling'' in Pisinger's article, randomly samples weights, $w_j$, from $\{1, 2, \dots ,1000\}$ and sets the profits as $v_j = 3\cdot \lceil w_j/3 \rceil$.

\subsubsection{Strongly Correlated Spanner Distribution}
Now we describe the first ``spanner'' distribution: ``Strong Spanner''.
In short, it samples $20$ elements from the strong distribution above, called the ``span'', then chooses $10$ random multiples from the span to form the knapsack. The elements chosen from the span are sampled with replacement. 
In full detail:
\begin{enumerate}
	\item Sample $20$ items from the strong knapsack distribution. Scale them: $w_j \gets \lceil 2w_j/3 \rceil$ and $v_j \gets \lceil 2v_j/3\rceil$.
			These scaled elements form the ``span'' and the span is a fixed list for the remainder of the sampling procedure.
	\item Repeat the following $10$ times: pick an element, $(w', v')$, from the spanner uniformly at random and scale this
			element with scaling, $s$, chosen uniformly at random from $\{1, 2, 3\}$. Add $(sw', sv')$ to
			the knapsack instance.
\end{enumerate}

\subsubsection{Profit Spanner Distribution}
The ``Profit Spanner'' distribution is the same as the previous, but the span is chosen from the profit distribution.
\begin{enumerate}
	\item Sample $20$ items from the profit knapsack distribution. Scale them: $w_j \gets \lceil 2w_j/3 \rceil$ and $v_j \gets \lceil 2v_j/3 \rceil$.
			These scaled elements form the ``span'' and the span is a fixed list for the remainder of the sampling procedure.
	\item Repeat the following $10$ times: pick an element, $(w', v')$, from the spanner uniformly at random and scale this
			element with scaling, $s$, chosen uniformly at random from $\{1, 2, 3\}$. Add $(sw', sv')$ to
			the knapsack instance.
\end{enumerate}

\subsection{Performance of Classical Algorithms on Hard KP Instances}
Figure \ref{fig:bars_results} and Table~\ref{table:alg_results_kp} in Appendix \ref{appendix:data} summarizes the performance of
four classical algorithms on the hard KP instances outlined above.
The probabilities and expectations in the table were estimated
by sampling 100 random instances for each distribution.
The details of our experimental choices for the classical
algorithms are as follows.
\paragraph{Greedy Algorithms (LG and VG)}
Both greedy algorithms are deterministic without additional parameters. 
Hence, we simply ran both greedy algorithms of Section~\ref{sec:classical_algorithms} on each of the random
KP instances and calculated the expectations and
probabilities in Table~\ref{table:alg_results_kp} in Appendix \ref{appendix:data}
by sample statistics.

\paragraph{Simulated Annealing (SA)}
The number of steps and the temperature are the main parameters for SA. 
For each KP instance, Simulated Annealing used $n=10$ steps and we optimized
the temperature for each input instance by sweeping a range
of temperatures.
Further, we gave simulated annealing a warm start by taking the starting state as the output of the Lazy Greedy algorithm.
On each input instance, the exact SA experiment
is as follows:
\begin{enumerate}
    \item Run Lazy Greedy on the instance to get a warm
            starting state.
    \item Sweep a range of temperature parameters and
            find an optimal temperature $T^*{>}0$ for the instance.
            Since simulated annealing algorithms
            are probabilistic, we did this by naive
            sampling: for each $T \in 100\ZZ \cap (0,2000]$,
            we ran simulated annealing ten times and used
            the $T^*$ with the best averaged score of those ten runs.
            We remark that choosing a finer grid than
            iterations of $100$ did not
            impact simulated annealing's performance.
        \item Run simulated annealing
            with $10$ steps from the warm start
            and with the
            temperature from the previous step.
            Record the highest score seen on the
            random walk and return it as output.
\end{enumerate}
\paragraph{Global Simulated Annealing}
On each KP instance, the GSA experiment ran the Lazy Greedy
algorithm for a warm start and optimized the temperature
for each instance using the same method as the SA experiment.
Then, the GSA experiment took the maximum score
of ten independent global steps (flip each bit
with probability $1/n$, one step in expectation)
from the warm start.
\paragraph{Analysis of Classical Performance}
Here we describe the results in Figure \ref{fig:bars_results} and Table~\ref{table:alg_results_kp} in Appendix \ref{appendix:data}
by considering each distribution by their ratios
$\zeta_i := v_i/w_i$. This
ratio determines how the greedy algorithms perform.
\begin{itemize}
    \item For the Strong distribution, $\zeta_i = 1/(1+1000/v_i) \in (0,1/2]$ increases
        directly with $v_i$. Therefore, Very Greedy 
        will \emph{always} match Lazy Greedy since the input
        is sorted by value $v$.
    \item For the Inverse Strong distribution, the values are similar to Strong
        but with a small variation: $\zeta_i = \frac{1}{1+(100+u_i)/v_i}$ where
        $u_i$ is uniformly random over $\{0, \pm 1, \pm 2\}$. Hence,
        we expect Very Greedy to outperform Lazy Greedy since this small variation
        will make the sorted list near the boundary, i.e.,
        an almost-full knapsack, unpredictable for the greedy algorithms. 
        In other words, looking ahead
        in the list is likely to increase the knapsack's value while staying under capacity.
    \item Something similar happens in the Profit distribution, though in
        a more structured manner. The ratios here are $\zeta_i = 1$ if $w_i \equiv 0 \mod 3$ and $1 + 3/w_i$ otherwise. In expectation,
        the last third of the inputs will be unsorted in terms of value
        and the first two-thirds are decreasing with weight.
        Therefore, we expect both greedy algorithms to perform
        poorly on these instances since a large fraction (roughly one third)
        of the input's ratios give no information on their value.
        For the other two-thirds, they will add the smallest,
        least valuable elements first.
    \item The spanner distributions are the same as the above except they sample with replacement, forming the span, then randomly scale the weights and values of those items in the span. 
    This random scaling will greatly affect the greedy algorithms, especially Lazy Greedy, since scaling has no effect on the $\zeta_i$'s but affects whether or not an item will fit and multiplies an element's value.
\end{itemize}

\subsection{Quantum Algorithmic Performance on Hard KP Instances}
Figure \ref{fig:bars_results} and Table~\ref{table:alg_results_kp} in Appendix \ref{appendix:data} summarize the performance of two quantum algorithms on the hard KP instances outlined in\S\ref{ssec:instances}.
The details on the quantum algorithms are as follows. 
\subsubsection{$\QKP$ Algorithms}
We simulate the Knapsack xQAOA algorithms described in Section~\ref{section:QAOAforKP} for depth $p{=}1$, using the Hourglass mixer $\Hourglass$ and the partitioned copula ring mixer $\tilde{U}^{\Copn}$. 
For both mixers, we run the corresponding xQAOA $n{=}10$ times, outputting the best solution that maximizes the objective function, assigning infeasible solutions to value $0$:
\begin{align}\label{eq:KP_sim-objective}
    f_{\text{obj}}(x) = 
    \begin{cases}
    v \cdot x & \text{if~} w {\cdot}x \leq c \\
    0 & \text{if~}  w{\cdot}x > c
    \end{cases}
\end{align}
where $a{\cdot}b$ denotes the dot product $\sum_{i=1}^n a_ib_i$.
With the definitions of the previous sections we are now ready to describe the optimization algorithms $\QKP_\Hourglass$ and $\QKP_\Cop$ that we propose, and simulate for solving Knapsack instances $(n,v,w,c)$. 
The only difference is that $\QKP_\Hourglass$ uses the Hourglass mixer (Sec.~\ref{sec:hourglass}), and $\QKP_\Cop$ uses the partitioned ring-Copula mixer $\tilde{U}^{\Copn}(\beta)$ (Sec.~\ref{sec:ring_copula}). 
In addition to the typical xQAOA rotation angles $\beta \in [0, \pi), \gamma \in [0, 2\pi)$, both algorithms have the parameter $k\in\RR^+$ to specify the bias strength, and the Copula mixer also has the correlation parameter $\theta\in[-1, +1]$.  

Note that $\beta, \gamma$ have different ranges since both mixing unitaries $U^{B_\Hourglass}(\beta)$ and $\tilde{U}^{\Cop}(\beta) $ are $\pi$-periodic (up to global phases), while the cost unitary $U^C(\gamma)$ is $2\pi$-periodic as $v_i \in \ZZ^+$.
 
The output states of the circuits have the following form.
\begin{align}
    \ket{\beta, \gamma, k}_{\Hourglass} &=
    U^{B_{\Hourglass(p(k))}}(\beta) U^C(\gamma) \ket{p(k)} \\
    \ket{\beta, \gamma, k, \theta}_\Cop &=
    \tilde{U}^{B_{\Cop(p(k), \theta)}}(\beta) U^C(\gamma) \ket{p(k)}
\end{align}
We prepare and measure this state $n$ times, outputting the best solution.
See Alg.~\ref{alg:QKP} for the full algorithms for $\QKP_\Hourglass$ and $\QKP_\Cop$.

\begingroup
\setlength\abovedisplayskip{2pt} 
\setlength\belowdisplayskip{0pt}
\SetKwInOut{Input}{Input}
\begin{algorithm}
\SetNoFillComment
\label{alg:QKP}
\caption{ $\QKP_\Hourglass \  /\  \QKP_\Cop (n, v, w, c, \gamma, \beta, k, *\theta)$ \ *($\theta$ is only required for $\Cop$)}
\SetAlgoLined
\Input{\underline{KP instance} $(n, v, w, c)$:\\
$n\in\ZZ^+, \ v, w \in \RR^n,\ c\in\RR^+$\\
\underline{Parameters} $(\gamma, \beta, k, *\theta)$:\\
$\gamma \in [0, \pi), \beta \in [0, 2\pi), \  k\in \RR^+, \  \theta\in [-1, +1]$\\
\tcp{ $\theta$ is only used for $\Cop$}
}
$p = (p_1, \dots , p_n) \gets \textsc{\textbf{Logistic}}(n, v, w, c, k)$ \tcp{Alg.~\ref{alg:logistic}}
\For{$t \gets 1$ \KwTo $n$ }{
    Initialize $n$ qubits as $\ket{p_1}\otimes\cdots\otimes \ket{p_n}$. \\
    Apply to the quantum register the energy function. 
    \begin{align}
    U^C(\gamma) &= U^{v_1}(\gamma)\otimes \cdots \otimes U^{v_n}(\gamma),
    \end{align} 
    where 
    $U^{v_i}(\gamma):\ket{x_i} \mapsto \e^{-\i \gamma v_i x_i}\ket{x_i}$. \\
    \If{ $\Hourglass$}{
        Apply to the register the Hourglass mixing unitary.
        \begin{align}
        U^{B_{\Hourglass}}(\beta) & =
        \e^{-\i\beta\Hourglass_{p_1}} 
        \otimes \cdots \otimes \e^{-\i\beta\Hourglass_{p_n}}
        \end{align}
    }
    \If{$\Cop$} {
        $(\theta_1,\dots,\theta_n)\gets \theta^n$ (setting all ${\theta}_i$ to $\theta$)\\
        Apply to the register the partitioned ring-copula mixing unitary $\tilde{U}^{\Cop}(\beta)$ (Eq.~\ref{eq:part_ring_copula_unitary}, Fig.~\ref{fig:ring_copula}).
        \begin{align}
            (\e^{-\i\beta\CopEn({p}, \theta)})(\e^{-\i\beta\CopOn({p}, \theta)})
        \end{align} 
    }
    Measure in the computational basis the $n$ bit outcome $z^{(t)}\in\{0,1\}^n$}
Determine which of the observed $z^{(1)},\dots,z^{(n)}$ maximizes the objective function $f_\text{obj}$ (Eq.~\ref{eq:KP_sim-objective}), and return that as the final output.   
\end{algorithm}
\endgroup

\begin{algorithm}
\label{alg:logistic}
\SetAlgoLined
\textbf{Input} : $n\in \ZZ^+,\  v, w\in\RR^n,\ c, k\in \RR^+$ \\
\textbf{Output} : $p=(p_1, p_2, \dots, p_n) \in [0, 1]^n$\\
 $r_i \gets v_i/w_i$ for all $i\in\{1,\dots,n\}$\\
    Calculate the LG solution and its stopping point $r_\mathrm{stop}$ \\
    With $C=|\sum_i w_i|/c-1$ and $(r_1,\dots,r_n,r_\mathrm{stop},k)$ use the logistic function of Equation~\ref{eq:logistic} to define the biases $p_1,\dots,p_n$\\
\caption{ \textsc{Logistic}$(n, v, w, c, k)$ }
\end{algorithm}

\subsubsection{Parameter Optimization}

Both of the Quantum algorithms we simulated $\QKP_\Hourglass$, and $\QKP_\Cop$ depend on the \xQAOA rotation angles $\gamma \in [0, 2\pi), \beta \in [0, \pi)$ as well as $k\in \RR^+$ which determines the bias strength as the steepness of the logistic curve \ifconference (Eq.~\ref{eq:logistic})\else(Fig.~\ref{fig:logistic})\fi. 
The Copula mixer also depends on the correlation parameter $\theta \in [-1, 1]$. In our simulations, we optimize each of these parameters \textit{per problem instance} to maximize the expectation value of $f_{\text{obj}}$ on the output of $\QKP$. 
For a  $\QKP_\Cop$ for a given problem instance $(n, v, w, c)$,  we optimize $\beta, \gamma$ for given $k, \theta$, by first doing a \textsc{GridSearch} (Algorithm~\ref{alg:grid_search}) of $N_\beta$ and $N_\gamma$ values of $\beta$ and $\gamma$ respectively. We use the best performing pair $\beta_0, \gamma_0$ that maximizes $\E[f_{\text{obj}}(\QKP_\Cop(n, v, w, c, \gamma_0, \beta_0))]$ as the initial state for the \text{BFGS} optimization algorithm.

To optimize parameters for the $\QKP_\Cop$ algorithm for a given problem instance $KP = (n, v, w, c)$, we try all combinations of $\theta{\in}\textsc{theta-range}$, $k{\in}\textsc{k-range}$ and determine the best corresponding $\beta$, $\gamma$ as described in the previous paragraph. We finally output the best $(\beta, \gamma), k, \theta$ to optimize $\E[f_\text{obj}[\QKP_\Cop(KP,  \gamma, \beta, k, \theta))]$ that was tried. 
Parameter optimization for $\QKP_\Hourglass$ is the same but without $\theta$.

We optimize over $k$-values,
$\textsc{k-range}{=}\{10, 11, \dots, 24 \}$, and \textsc{GridSearch} with $N_\gamma{=}N_\beta{=}50$ different equally spaced values for $\gamma, \beta$. 
We use $\theta$-values $\textsc{theta-range} = \{0, -\frac{1}{2}, -1 \}$ to see uncorrelated,  ``half anti-correlated'', and maximally anti-correlated.

Note that $\QKP$ outputs the best of $n$ samples from our quantum circuit. Therefore, calculating $\E[f_\text{obj}(\QKP(\cdot))]$ is the expected best of $n$ samples. Moreover, we explicitly calculated this expectation value since our classical simulation kept track of the entire statevector. However, this is much less reasonable for classical simulation of larger instances (or running this on a quantum circuit), where we instead  would approximate this expectation value. 

\begin{algorithm}
\SetAlgoLined
\For{$k \gets \textsc{k-range}$} {
    \For{$\theta \gets \textsc{theta-range}$} {
        $\beta_0, \gamma_0 \gets \mathbf{\beta\gamma}$-$\textsc{GridSearch}(\QKP, N_\beta, N_\gamma)$\\
        $((\beta, \gamma)_\textsf{bfgs}, \textit{val}_\textsf{bfgs}, x_\textsf{bfgs}) \gets \textsc{BFGS}(\E[f_\text{obj}(\QKP)], (\beta_0, \gamma_0))$\\
        \If{$\textit{val}_\textsf{bfgs} > \textit{val}^*$} {
            $\textit{val}^* \gets \textit{val}_\textsf{bfgs}$\\
            $(\beta^*, \gamma^*) \gets (\beta, \gamma)_\textsf{bfgs}$\\
            $x^* \gets x_\textsf{bfgs}$
        }
    }
}
\Return $(\beta^*, \gamma^*), x^*, \textit{val}^*$
\caption{Parameter Optimization}\label{alg:param_optimization}
\end{algorithm}
\vspace{-\topsep}
\begin{algorithm}
\SetAlgoLined
\textbf{Input} : $\QKP \in \{ \QKP_\Hourglass, \QKP_{\Hourglass_\Cop}\}$ (including parameters),  $N_\beta, N_\gamma \in \ZZ^+$, specifying grid search for $\beta \in [0, \pi)], \gamma \in [0, 2\pi).$\\
\textbf{Output} : $\beta^*\in [0, \pi), \gamma^* \in [0, 2\pi)$ Best in grid.\\
\For{$\beta \gets \{0, \pi\frac{1}{N_\beta}, \frac{\pi\cdot 2}{N_\beta}, \dots, \frac{\pi\cdot (N_\beta-1)}{N_\beta} \}$} {
    \For{$\gamma \gets \{0, 2\pi\frac{1}{N_\gamma}, \frac{2\pi\cdot 2}{N_\gamma}, \dots, \frac{2\pi\cdot (N_\gamma-1)}{N_\gamma}$ \}} {
        $\textit{val} \gets \E[f_\text{obj}(\QKP(\gamma, \beta))]$\\
        \If{$\textit{val} > \textit{val}^*$} {
            $\textit{val}^* \gets \textit{val}$\\
            $\beta^* \gets \beta, \ \gamma^* \gets \gamma$
        }
    }
}
\Return{$\beta^*, \gamma^*$}
\caption{$\mathbf{\beta\gamma}$-$\textsc{GridSearch}(\QKP, N_\beta, N_\gamma)$:\  $(\beta, \gamma)$ Brute-force Optimization}\label{alg:grid_search}
\end{algorithm}

\subsubsection{Simulation Results}
In Figure \ref{fig:bars_results} below and Table~\ref{table:alg_results_kp} in Appendix \ref{appendix:data} we show various performance statistics on the different classical and quantum Knapsack algorithms applied to 100 instances sampled from each of the hard problem distributions described in \S\ref{ssec:instances}. Although $\QKP_\Hourglass$ does not use entangling gates, it performs surprisingly well compared to the classical algorithms. The Copula mixer further improves performance as $\QKP_\Cop$ outperforms $\QKP_\Hourglass$.
Note that the analysis of the quantum algorithm does not assume post-selection as infeasible solutions are treated as solutions with $f_\text{obj}  = 0$; see Eq.~\ref{eq:KP_sim-objective}.

Anti-correlating neighboring bits more with the Copula mixer, typically resulted in a better solution. Figure~\ref{fig:k_trends_FGM} depicts how the average approximation ratio $\E[f_{obj}(x)/f_{obj}^*]$ depends on $k, \theta$ for each problem class.


\pgfplotstableread[row sep=\\,col sep=&]{
    Algorithm & strong & invstrong & profit & sspanner & pspanner \\
    LG        & 0.11   & 0.15      & 0.02   & 0.09     & 0.02\\
    VG        & 0.11   & 0.46      & 0.07   & 0.10     & 0.08\\
    SA        & 0.146  & 0.272     & 0.061  & 0.141    & 0.086\\
    GSA       & 0.133  & 0.253     & 0.047  & 0.122    & 0.058\\
    HG        & 0.470  & 0.549     & 0.088  & 0.440    & 0.106\\
    Cop       & 0.478  & 0.580     & 0.121  & 0.473    & 0.129\\
    X         & 0.355  & 0.080     & 0.090  & 0.156    & 0.076 \\
}\proboptTable

\pgfplotstableread[row sep=\\,col sep=&]{
    Algorithm & strong & invstrong & profit & sspanner & pspanner \\
    VG  & 0.00  & 0.79  & 0.72  & 0.31  & 0.76 \\
    SA  & 0.462 & 0.785 & 0.768 & 0.533 & 0.800 \\
    GSA & 0.283 & 0.589 & 0.565 & 0.369 & 0.615 \\
    HG  & 0.838 & 0.814 & 0.896 & 0.842 & 0.867  \\ 
    Cop & 0.844 & 0.820 & 0.926 & 0.849 & 0.882 \\
    X   & 0.782 & 0.714 & 0.902 & 0.758 & 0.858 \\
}\probbeatLGTable

\pgfplotstableread[row sep=\\,col sep=&]{
    Algorithm & strong & invstrong & profit & sspanner & pspanner \\
    SA  & 0.462 & 0.045 & 0.297 & 0.355 & 0.291 \\
    GSA & 0.283 & 0.014 & 0.142 & 0.223 & 0.143 \\
    HG  & 0.838 & 0.421 & 0.631 & 0.804 & 0.683 \\
    Cop & 0.844 & 0.435 & 0.681 & 0.813 & 0.705    \\
    X   & 0.782 & 0.190 & 0.616 & 0.703 & 0.674 \\
}\probbeatVGTable

\pgfplotstableread[row sep=\\,col sep=&]{
    Algorithm & strong & invstrong & profit & sspanner & pspanner \\
    LG  & 0.905 & 0.873 & 0.840 & 0.863 & 0.802 \\
    VG  & 0.905 & 0.985 & 0.952 & 0.916 & 0.958 \\
    SA  & 0.945 & 0.965 & 0.951 & 0.935 & 0.943 \\
    GSA & 0.928 & 0.948 & 0.917 & 0.913 & 0.901 \\
    HG  & 0.986 & 0.990 & 0.972 & 0.983 & 0.975 \\
    Cop & 0.988 & 0.993 & 0.979 & 0.986 & 0.979 \\
    X   & 0.974 & 0.947 & 0.973 & 0.957 & 0.970 \\
}\expectedratioTable

\pgfplotstabletranspose[string type,
                        colnames from=Algorithm,
                        input colnames to=Algorithm
                        ]\transproboptTable{\proboptTable}
                        
\pgfplotstabletranspose[string type,
                        colnames from=Algorithm,
                        input colnames to=Algorithm
                        ]\transprobbeatLGTable{\probbeatLGTable}
                        
\pgfplotstabletranspose[string type,
                        colnames from=Algorithm,
                        input colnames to=Algorithm
                        ]\transprobbeatVGTable{\probbeatVGTable}                        
\pgfplotstabletranspose[string type,
                        colnames from=Algorithm,
                        input colnames to=Algorithm
                        ]\transexpectedratioTable{\expectedratioTable}        
                        
\begin{figure}[p]
\begin{tikzpicture}
\begin{axis}[
        ybar=6pt,
        bar width=8pt,
        title=Probability of Optimality,
        symbolic x coords={strong, invstrong, profit, sspanner, pspanner},
        ybar, axis on top,
        xtick distance=1,
        clip=false,
        height=0.3\linewidth, width=.95\linewidth,
        ymajorgrids, tick align=inside,
        enlarge y limits={value=.1,upper},
        axis x line*=bottom,
        axis y line*=right,
        y axis line style={opacity=0},
        legend style={
            at={(1.15, 0.5)},
            anchor=center,
            /tikz/every even column/.append style={column sep=0.5cm}
        },
    ]
    \addplot[fill=LGcolor]  table[x=Algorithm,y=LG]{\transproboptTable};
    \addplot[fill=VGcolor]  table[x=Algorithm,y=VG]{\transproboptTable};
    \addplot[fill=SAcolor]  table[x=Algorithm,y=SA]{\transproboptTable};
    \addplot[fill=GSAcolor] table[x=Algorithm,y=GSA]{\transproboptTable};
    \addplot[fill=Xcolor] table[x=Algorithm,y=X]{\transproboptTable};
    \addplot[fill=HGcolor]  table[x=Algorithm,y=HG]{\transproboptTable};
    \addplot[fill=Copcolor] table[x=Algorithm,y=Cop]{\transproboptTable};

    \legend{LG, VG, SA, GSA, $\QKP_\Xgate$, $\QKP_\Hourglass$, $\QKP_\Cop$}
    \end{axis}
\end{tikzpicture}
\newline

\begin{tikzpicture}
\begin{axis}[
        ybar=6pt,
        bar width=8pt,
        title=Expectation of Approximation Ratio,
        symbolic x coords={strong, invstrong, profit, sspanner, pspanner},
        ybar, axis on top,
        xtick distance=1,
        clip=false,
        height=0.3\linewidth, width=.95\linewidth,
        ymajorgrids, tick align=inside,
        enlarge y limits={value=.1,upper},
        axis x line*=bottom,
        axis y line*=right,
        y axis line style={opacity=0},
        legend style={
            at={(1.15, 0.5)},
            anchor=center,
            /tikz/every even column/.append style={column sep=0.5cm}
        },
    ]
    \addplot[fill=LGcolor]  table[x=Algorithm,y=LG]{\transexpectedratioTable};
    \addplot[fill=VGcolor]  table[x=Algorithm,y=VG]{\transexpectedratioTable};
    \addplot[fill=SAcolor]  table[x=Algorithm,y=SA]{\transexpectedratioTable};
    \addplot[fill=GSAcolor] table[x=Algorithm,y=GSA]{\transexpectedratioTable};
    \addplot[fill=Xcolor] table[x=Algorithm,y=X]{\transexpectedratioTable};
    \addplot[fill=HGcolor]  table[x=Algorithm,y=HG]{\transexpectedratioTable};
    \addplot[fill=Copcolor] table[x=Algorithm,y=Cop]{\transexpectedratioTable};

    \legend{LG, VG, SA, GSA,$\QKP_\Xgate$, $\QKP_\Hourglass$, $\QKP_\Cop$}
    \end{axis}
\end{tikzpicture}
\newline

\begin{tikzpicture}
\begin{axis}[
        ybar=6pt,
        bar width=8pt,
        title=Probability of Outperforming Lazy Greedy,
        symbolic x coords={strong, invstrong, profit, sspanner, pspanner},
        ybar, axis on top,
        xtick distance=1,
        clip=false,
        height=0.3\linewidth, width=.95\linewidth,
        ymajorgrids, tick align=inside,
        enlarge y limits={value=.1,upper},
        axis x line*=bottom,
        axis y line*=right,
        y axis line style={opacity=0},
        legend style={
            at={(1.15,.5)},
            anchor=center,
            /tikz/every even column/.append style={column sep=0.5cm}
        },
    ]

    \addplot[fill=VGcolor]  table[x=Algorithm,y=VG]{\transprobbeatLGTable};
    \addplot[fill=SAcolor]  table[x=Algorithm,y=SA]{\transprobbeatLGTable};
    \addplot[fill=GSAcolor] table[x=Algorithm,y=GSA]{\transprobbeatLGTable};
    \addplot[fill=Xcolor] table[x=Algorithm,y=X]{\transprobbeatLGTable};
    \addplot[fill=HGcolor]  table[x=Algorithm,y=HG]{\transprobbeatLGTable};
    \addplot[fill=Copcolor] table[x=Algorithm,y=Cop]{\transprobbeatLGTable};

    \legend{VG, SA, GSA, $\QKP_\Xgate$, $\QKP_\Hourglass$, $\QKP_\Cop$}
    \end{axis}
\end{tikzpicture}
\newline

\begin{tikzpicture}
\begin{axis}[
        ybar=6pt,
        bar width=8pt,
        title=Probability of Outperforming Very Greedy,
        symbolic x coords={strong, invstrong, profit, sspanner, pspanner},
        ybar, axis on top,
        xtick distance=1,
        clip=false,
        height=0.3\linewidth, width=.95\linewidth,
        ymajorgrids, tick align=inside,
        enlarge y limits={value=.1,upper},
        axis x line*=bottom,
        axis y line*=right,
        y axis line style={opacity=0},
        legend style={
            at={(1.15, 0.5)},
            anchor=center,
            /tikz/every even column/.append style={column sep=0.5cm}
        },
    ]
    \addplot[fill=SAcolor]  table[x=Algorithm,y=SA]{\transprobbeatVGTable};
    \addplot[fill=GSAcolor] table[x=Algorithm,y=GSA]{\transprobbeatVGTable};
    \addplot[fill=Xcolor] table[x=Algorithm,y=X]{\transprobbeatVGTable};
    \addplot[fill=HGcolor]  table[x=Algorithm,y=HG]{\transprobbeatVGTable};
    \addplot[fill=Copcolor] table[x=Algorithm,y=Cop]{\transprobbeatVGTable};

    \legend{SA, GSA,$\QKP_\Xgate$, $\QKP_\Hourglass$, $\QKP_\Cop$}
    \end{axis}
\end{tikzpicture}

\caption{\label{fig:bars_results} Visualization of each algorithm's performance for each distribution of Knapsack Problem instances. See  Appendix \ref{appendix:data} and Table \ref{table:alg_results_kp}. for full details.}
\end{figure}


\subsubsection{Parameter Sensitivity}\label{ssec:param_sensitivity}
\begin{figure}
    \centering
    \includegraphics[width=0.8\columnwidth]{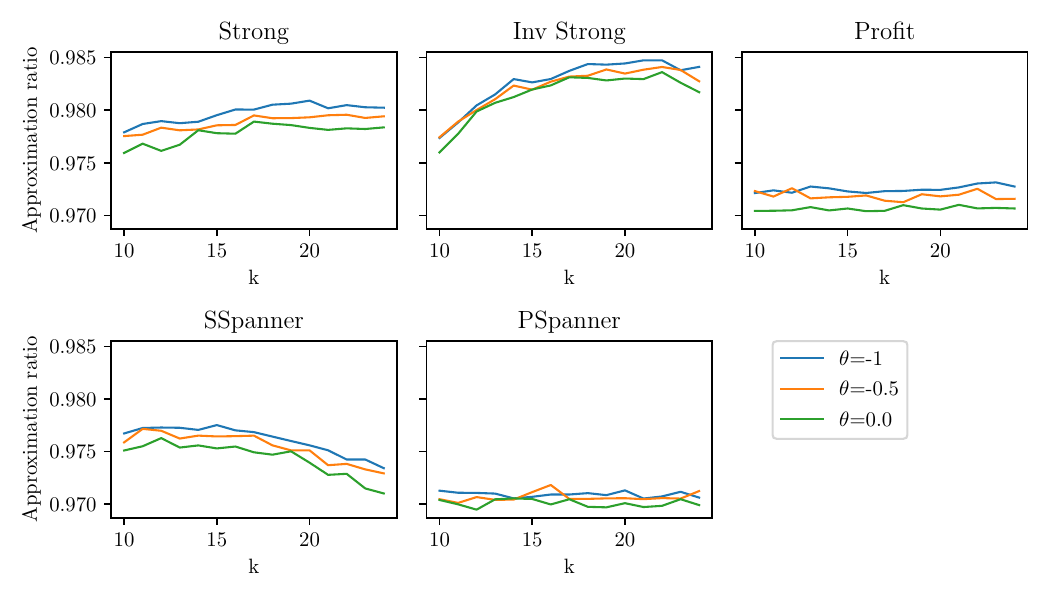}
    \caption{Average approximation ratio as a function of $k$ for each $\theta$ of the biased xQAOA with ring copula mixer for Knapsack Problem. Each subplot averages over the approximation ratio for 100 samples from a specific hard instance distribution described in Section~\ref{ssec:instances}. $\theta{=}0$ corresponds to an \textit{uncorrelated} mixer, which is equivalent to the Hourglass mixer $B_\Hourglass$, whereas $\theta{=}1$ is maximally anti-correlated.}
    \label{fig:k_trends_FGM}
\end{figure}

\begin{figure}[H]
    \centering
    \includegraphics[width=\columnwidth]{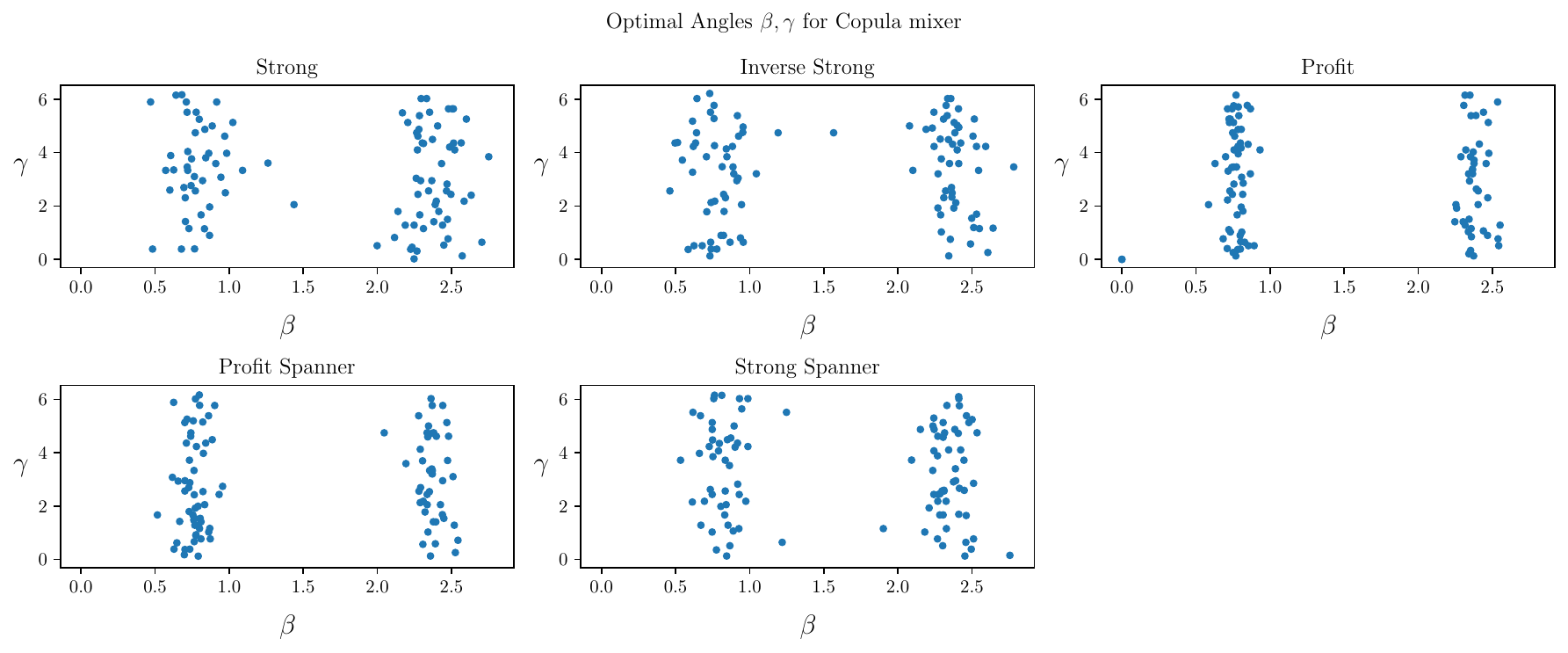}
    \caption{Optimal angles $\beta, \gamma$ for $\QKP_\Cop$. Each dot corresponds to a single problem instance. $\beta, \gamma$ are optimal in that they maximize the expected value of $f_\text{obj}$ of the output of Algorithm \ref{alg:QKP}.}
    \label{fig:amh_cop_opt_angles}
\end{figure}

\begin{figure}[H]
    \centering
    \includegraphics[width=\columnwidth]{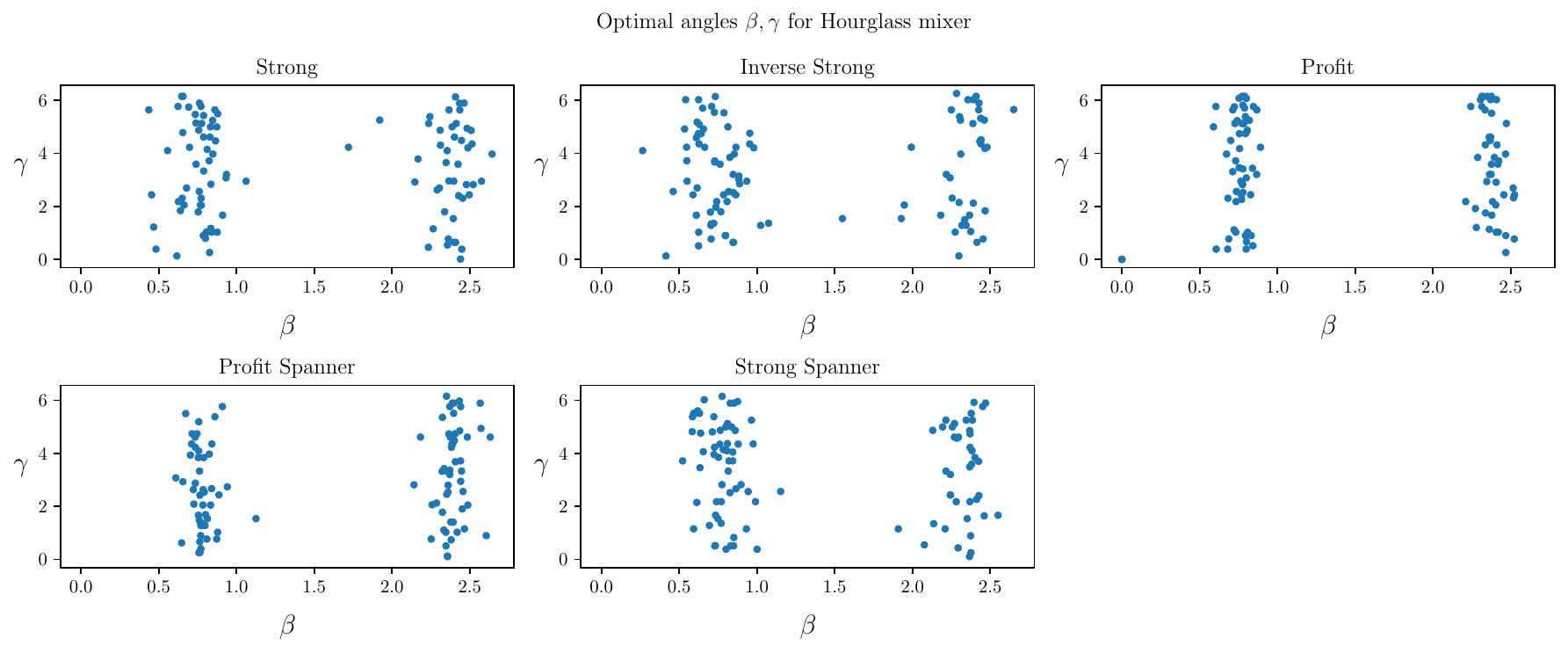}
    \caption{Optimal angles $\beta, \gamma$ for $\QKP_\Hourglass$. Each dot corresponds to a single problem instance. $\beta, \gamma$ are optimal in that they maximize the expected value of $f_\text{obj}$ of the output of Algorithm \ref{alg:QKP}.}
    \label{fig:hg_opt_angles}
\end{figure}

It is sensible for the reader at this point to wonder if the performance of the algorithms discussed relies on heavy optimization of the parameters $\beta, \gamma, k, \theta$, and if such parameter optimization will scale with problem size. While the question of scaling remains to be explored by simulating larger instances, we observe numerically that on these instances of size $n=10$, the algorithms still perform well after fixing some parameters.
As shown in Figure \ref{fig:k_trends_FGM}, $\QKP_\Cop$ typically performs better when we maximally anti-correlate the neighboring bits $\theta = -1$. Moreover, Figure \ref{fig:k_trends_FGM}, highlights that the performance is not very sensitive to the value of $k$. In Figures \ref{fig:amh_cop_opt_angles} and \ref{fig:hg_opt_angles} we observe that for both the Hourglass and the Copula algorithms, the optimal value for $\beta$ concentrates around $\pi/4$ and $3\pi/4$.


\section{Conclusions and Future Work}
\label{sec:concusion}
The results in this paper show how the presented constant depth quantum optimization heuristics perform distinctly better than the similarly shallow classical greedy and simulated annealing algorithms. 
While it is likely that more elaborate classical algorithms will outperform our quantum algorithms, we argue that the presented comparison is a fair one as it involves quantum and classical algorithms of equal simplicity. In Appendix \ref{appendix:data}, we include scatter plots that compare the performance of these algorithms for each individual problem instance.
We note that the dependence on the choice of parameters is relatively weak, i.e., it is possible to maintain good performance by using instance-independent parameters (see \S \ref{ssec:param_sensitivity} and Figure~\ref{fig:k_trends_FGM}). 

For future work, we propose to study how the presented quantum algorithms perform on larger problem instances ($n{\gg}10$) and how to increase the performance by using deeper circuits ($p{\gg}1$). 
Finally, the general quantum techniques of using biased initial states and mixers that preserve this bias should have applications for other constrained optimization problems as well.  
It is our hope that doing so will greatly expand the promise of quantum optimization heuristics for problems of practical importance. 

\paragraph*{Acknowledgement} The research presented in this article was made possible by a SBIR/STTR grant of the Department of Defense; contract no.\ FA8750-20-P-1711.

%% file: appendix.tex
\appendix
\section{Data}\label{appendix:data}

\begin{table} [ht!]
\begin{centering}
\small
\begin{tabular}{c|S[table-format=0.3] S[table-format=0.3] S[table-format=0.3] S[table-format=0.3] S[table-format=0.3]}
 &          & {\textbf{Inv}} &  & {\textbf{Strong}} & {\textbf{Profit}} \\
 & {\textbf{Strong}} & {\textbf{Strong}} & {\textbf{Profit}} & {\textbf{Spanner}} & {\textbf{Spanner}} \\
 \\
& \multicolumn{5}{c}{\emph{Probability of Optimality}}\\
\hline
 \LG  & 0.11& 0.15 & 0.02 & 0.09 & 0.02 \\
 \VG  & 0.11 & 0.46 & 0.07 & 0.10 & 0.08 \\
 \SA & 0.146 & 0.272 & 0.061 & 0.141 & 0.086 \\
\GSA & 0.133 & 0.253 & 0.047 & 0.122 & 0.058 \\
$\QKP_{\Xgate}$  & 0.355  & 0.080     & 0.090  & 0.156    & 0.076 \\
\rowcolor{orange} $\QKP_\Hourglass$ & 0.470 & 0.549 & 0.088 & 0.440 & 0.106
\\
\rowcolor{yellow}$\QKP_\Cop$& 0.478 & 0.580 & 0.121 & 0.473 & 0.129\\
\hline\\
& \multicolumn{5}{c}{\emph{Probability of Outperforming Lazy Greedy}}  \\\hline
 \VG & 0.00 & 0.79 & 0.72 & 0.31 & 0.76 \\
 \SA & 0.462 & 0.785 & 0.768 & 0.533 & 0.800 \\
 \GSA & 0.283 & 0.589 & 0.565 & 0.369 & 0.615 \\
 $\QKP_{\Xgate}$   & 0.782 & 0.714 & 0.902 & 0.758 & 0.858 \\
\rowcolor{orange} $\QKP_\Hourglass$  & 0.838 & 0.814 & 0.896 & 0.842 & 0.867  \\ 
\rowcolor{yellow} $\QKP_\Cop$ & 0.844 & 0.820 & 0.926 & 0.849 & 0.882   \\\hline
\\
& \multicolumn{5}{c}{\emph{Probability of Outperforming Very Greedy}}  \\\hline
 \SA & 0.462 & 0.045 & 0.297 & 0.355 & 0.291 \\
 \GSA & 0.283 & 0.014 & 0.142 & 0.223 & 0.143 \\
 $\QKP_{\Xgate}$   & 0.782 & 0.190 & 0.616 & 0.703 & 0.674 \\
\rowcolor{orange} $\QKP_\Hourglass$ &  0.838 &  0.421 &  0.631 &  0.804 & 0.683 \\
\rowcolor{yellow}$\QKP_\Cop$& 0.844 & 0.435 & 0.681 & 0.813 & 0.705    \\\hline \\
& \multicolumn{5}{c}{\emph{Expected Approximation Ratio}}  \\\hline
 \LG     & 0.905 & 0.873 & 0.840 & 0.863 & 0.802 \\
 \VG        & 0.905 & 0.985 & 0.952 & 0.916 & 0.958 \\
 \SA        & 0.945 & 0.965 & 0.951 & 0.935 & 0.943 \\
\GSA       & 0.928 & 0.948 & 0.917 & 0.913 & 0.901 \\
 $\QKP_{\Xgate}$   & 0.974 & 0.947 & 0.973 & 0.957 & 0.970 \\
\rowcolor{orange} $\QKP_\Hourglass$ & 0.986 & 0.990 & 0.972 & 0.983 & 0.975 \\
\rowcolor{yellow} $\QKP_\Cop$ & 0.988 & 0.993 & 0.979 & 0.986 & 0.979 \\\hline
 \end{tabular}
\end{centering}
\caption{Summary of results for the algorithms analyzed in this paper. The four classical optimization heuristics are: Lazy Greedy (LG), Very Greedy (VG), Simulated Annealing (SA), and Global Simulated Annealing (GSA). 
Our quantum optimization heuristics are the standard QAOA with the $\Xgate$ mixer, denoted $\QKP_\Xgate$, the xQAOA+Hourglass algorithm $\QKP_\Hourglass$ (in orange rows) and the xQAOA+Copula algorithm $\QKP_\Cop$ (in yellow rows) as defined in Algorithm~\ref{alg:QKP}.
The $5$ hard Knapsack scenarios (Strong, Inv Strong, Profit, Strong Spanner, Profit Spanner, as described in Section~\ref{ssec:instances}) are listed in the $5$ different columns. 
For each scenario we generated $100$ random instances with problem size $n{=}10$, yielding a total of $500$ problem instances. 
For each solver and scenario, the top table lists the probability that the algorithm finds the perfect solution of a uniformly sampled problem instance from the scenario. Hence we see, for example, that the deterministic VG algorithm perfectly solves $7$ out of the $100$ instances of the Profit scenario. 
For the solvers \SA, \GSA, $\QKP_\Hourglass$, $\QKP_\Cop$ the Probability of Optimality takes into account the probabilistic nature of these solvers. 
Similarly, an entry of the second (third) table shows for a given scenario and solver, the probability that the solver outputs an answer that is strictly better than the answer of the deterministic Lazy Greedy (Very Greedy) on a uniformly sampled problem instance from the scenario. Hence we see that Very Greedy is never better than Lazy Greedy on the instances from the Strong scenario, but also that Very Greedy is better than Lazy Greedy on $76$ of the $100$ instances from Profit Spanner.
Each entry in the bottom table shows for a solver and scenario the expected approximation ratio for a uniformly sampled problem instance, where the ratio is the expectation of the output value, divided by the optimal value. If the solver outputs an invalid string we treat this as a solution with value $0$; hence these expectations are unconditional ones.
As the ratios are close to $1$ we see that the solvers almost always produce valid outputs.
Throughout the table, larger values are better and we thus see how for all five scenarios and four measures of success, the quantum heuristics outperform the classical ones.
}
 \label{table:alg_results_kp}
\end{table}

\begin{figure}[h]
    \centering
    \includegraphics[width=\columnwidth]{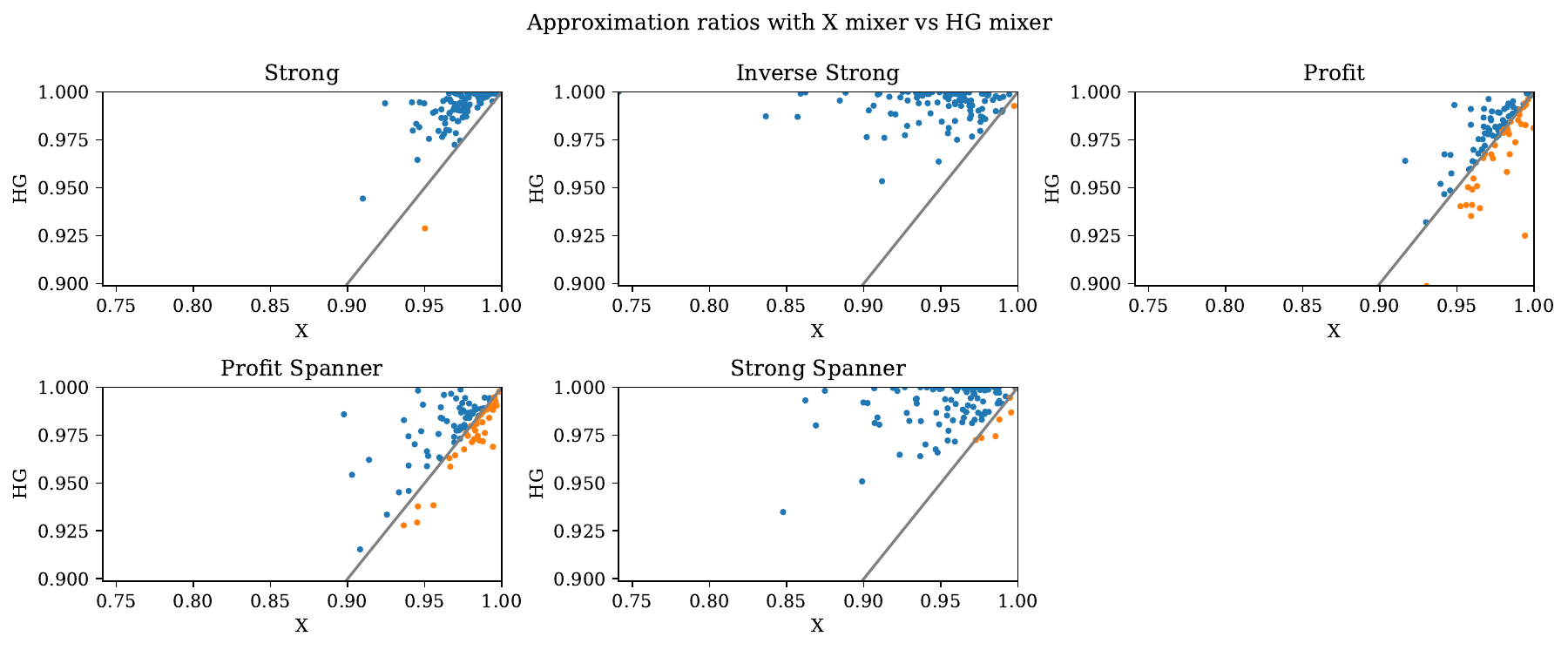}
    \caption{Approximation ratios of $\QKP_\Hourglass$ and $\QKP_\Xgate$ for each problem instance. Each dot corresponds to a problem instance. A dot $i$ is colored blue if $\E[\QKP_\Hourglass(i)] > \E[\QKP_\Xgate(i)]$, and orange otherwise.}
    \label{fig:hg_vs_x_ratios}
\end{figure}
\begin{figure}
    \centering
    \includegraphics[width=\columnwidth]{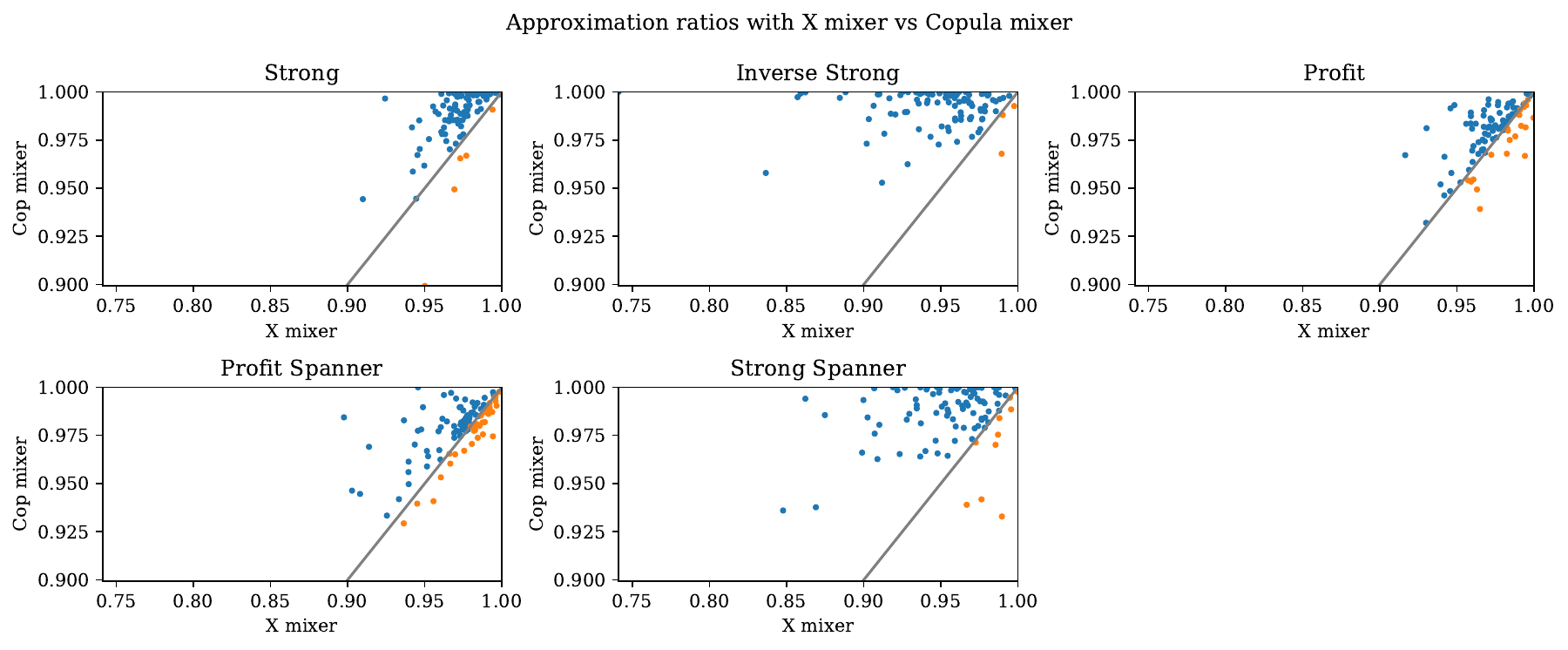}
    \caption{Approximation ratios of $\QKP_\Cop$ and $\QKP_\Xgate$ for each problem instance. Each dot corresponds to a problem instance. A dot $i$ is colored blue if $\E[\QKP_\Cop(i)] > \E[\QKP_\Xgate(i)]$, and orange otherwise.}
    \label{fig:cop_vs_x_ratios}
\end{figure}
\begin{figure}
    \centering
    \includegraphics[width=\columnwidth]{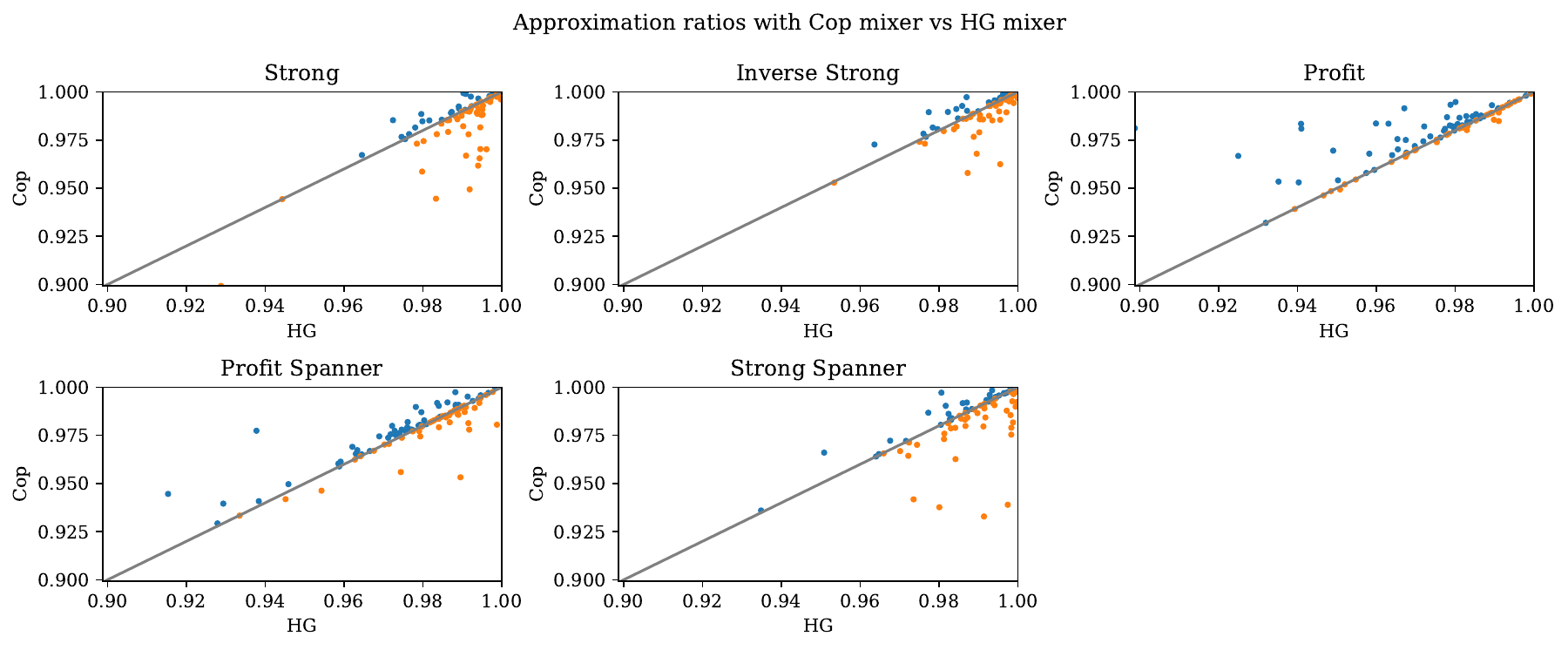}
    \caption{Approximation ratios of $\QKP_\Cop$ and $\QKP_\Hourglass$ for each problem instance. Each dot corresponds to a problem instance. A dot $i$ is colored blue if $\E[\QKP_\Cop(i)] > \E[\QKP_\Hourglass(i)]$, and orange otherwise.}
    \label{fig:cop_vs_hg_ratios}
\end{figure}